\documentclass[aps,prb,twocolumn,keywords,nopacs,amssymb]{revtex4}
\pdfoutput=1

\usepackage{graphicx}
\usepackage{dcolumn}
\usepackage{bm}

\def\be{\begin{equation}}
\def\en{\end{equation}} 
\def\p{\partial }  
\def\ve{\varepsilon}
\def\gs{\gtrsim}
\def\ls{\lesssim}

\def\bea{\begin{eqnarray}}
\def\ena{\end{eqnarray}}

\def\ns{n_{\rm MOH}} 
\def\nM{n_{\rm M}} 
\def\nO{n_{\rm OH}} 
\def\nH{n_{\rm H}}

\def\ms{\mu_{\rm MOH}} 
\def\mM{\mu_{\rm M}} 
\def\mO{\mu_{\rm OH}} 
\def\mH{\mu_{\rm H}}

\def\lsa{\lambda_{\rm MOH}} 
\def\lM{\lambda_{\rm M}} 
\def\lO{\lambda_{\rm OH}} 
\def\lH{\lambda_{\rm H}}

\begin{document}
\preprint{APS}
\title{Ionization at a solid-water interface in an applied electric field: 
Charge regulation}

\author{Ryuichi Okamoto$^1$ and Akira Onuki$^2$}
\affiliation{
$^1$Department of chemistry, 
Tokyo Metropolitan University, Hachioji,
Tokyo 192-0397,  Japan\\
$^2$Department of Physics, Kyoto University, Kyoto 606-8502, Japan}
\date{\today}

\date{\today}

\begin{abstract}
We investigate ionization at a solid-water interface in applied electric field. We attach an electrode to   a  dielectric    film   
bearing silanol or   carboxyl   groups with an areal density $\Gamma_0$, where 
the degree of dissociation 
 $\alpha$  is   determined by  the proton density in water 
 close to the film.
We show how $\alpha$ depends on the density $n_0$  of NaOH 
in water and  the surface charge density 
 $\sigma_m$ on the electrode.  For $\sigma_m>0$,  the protons are expelled 
away from the film, leading to  an increase in $\alpha$. 
In particular, in the range  $0<\sigma_m<e\Gamma_0$, 
self-regulation occurs to realize   $\alpha \cong \sigma_m/e\Gamma_0 $  for 
 $n_0\ll n_c$, where  $n_c$  is $0.01$ mol$/$L for silica 
surfaces and  is  $2\times 10^{-5}$ mol$/$L for carboxyl-bearing surfaces.  
We also examine the charge regulation  
with decreasing the cell thickness $H$ below the Debye length 
$\kappa^{-1}$, where a crossover occurs  at the Gouy-Chapman length. 
In particular, when  $\sigma_m \sim e\Gamma_0$ and   $H\ll \kappa^{-1}$,  
the surface charges remain  
 only partially screened by ions, 
leading to  an electric field in the interior. 
\end{abstract}

\maketitle

%
%

\section{Introduction}

Numerous papers have been written on 
 various aspects of the electric double layer 
 at a solid-water interface 
\cite{Hunter,Is,Butt,Andel3}. However, not enough 
attention has yet been paid on  physics and 
chemistry of ionizable solid surfaces 
 in contact with  an aqueous electrolyte 
solution.  As a well-known  example 
 \cite{Andel3,Hunter,Is,Butt,Sonn,Hal,Hie,Wes,Fuer,Leckie},  
 silanol  groups SiOH 
on a silica oxide  surface  at a density $\Gamma_0$ 
dissociate  into SiO$^-$ and mobile protons H$^+$. 
The surface charge density 
is   $-e\Gamma_0\alpha $, 
where $\alpha$ is   the degree of dissociation.  In this case, 
$\alpha$  is   determined by  the proton density  $\nH(0)$ in water 
immediately close to the surface.
This  $\nH(0)$  is different from the bulk proton density $\nH^0$,  
depending on the solution composition. 
Also  when  two ionizable  surfaces approach, 
 $\alpha$ and  the force between them change at small  separation 
\cite{Ga,Is1,Is,Zhao,BehR,Par,Chan}. 
These behaviors are  often referred to as  {\it charge regulation}. 
 We note that similar phenomena  with variable charges are  
ubiquitous in soft matters.

We mention   theoretical papers  on the charge regulation 
on solid-water interfaces with salts 
\cite{Par,Beh,Beh1,Beh0,Lowen,Chan,BehR,AndelEPL,Bie}.  
 Ninham and Parsegian\cite{Par}  first presented a 
model of surface ionization using the  mass action law 
as the boundary condition of the Poisson-Boltzmann (PB) equation, 
where $\Gamma_0$ and the dissociation 
constant $K_s$ are relevant parameters. 
They  determined  $\alpha$,  the charge distribution, and 
the electric potential  self-consistently.
Behrens {\it et al.}\cite{Beh,Beh0,Beh1} 
solved these equations  accounting  for 
a potential drop across the Stern layer formed on  
a solid surface in  water.
These theories are based on 
one-dimensional (1D)  calculations of the PB 
solutions  for symmetric 
ionizable surfaces.   Behrens {\it et al.}
  also devised a formula for  ionization of large  colloidal  
particles. 

In this paper, we   construct  a free energy 
functional for the ion densities $n_i$ and $\alpha$ 
in applied  field, accounting for the chemical reactions 
in the bulk and on the surface. 
We then  examine the effect of  sodium hydroxide NaOH   
at a low density $n_0$, where  the hydroxyl 
density  $n_{\rm OH}$ 
and  the proton density $\nH$  are related by 
the dissociation law $\nO \nH=10^{-14}$ mol$^2/$L$^2$  
 in bulk  water  \cite{Eigen}.  
Thus,   adding  NaOH  serves to decrease 
 $\nH(0)$  and increase $\alpha$.  
In experiments,   the surface ionization due to deprotonation  
 increased  with  addition of OH$^-$  (with increasing  pH) 
 \cite{Gisler,Yama,Yama1,Beh4,BehR}. 
We   also apply an electric field to the system 
 attaching a planar electrode with a surface charge density 
$\sigma_m$.  Then, there arises  another kind of charge regulation. 
We assume that 
the electric double layer next to the film is determined 
by the effective  density 
 $\sigma_{\rm eff}= -e\Gamma_0\alpha+\sigma_m$. 
Thus, if    $\sigma_m$ is positive (negative), the cations 
including the protons tend to be  
repelled from (attracted to) the film, leading to 
   an increase (a decrease) in $\alpha$.  
We have  $\alpha \to 0$ for $\sigma_m<0$ 
and $\alpha \to 1$ for $\sigma_m>e\Gamma_0$. 
However,   in the range  $0<\sigma_m<e\Gamma_0$, 
  marked  {\it self-regulation} behavior emeges, where  
 $\alpha$ approaches $\sigma_m/e\Gamma_0$ and $\sigma_{\rm eff}$ 
 nearly vanishes.  In this effect, 
 the NaOH density $n_0$ needs to be 
 smaller than a characteristic density $n_c$, where  
  $n_c$ is  much larger than the hydroxyl 
density  $ 10^{-7}$ mol$/$L  in pure water. 
For small   wall separation $H$ and for not small $\sigma_m$, 
 screening of the surface charges can  only be partial,  
leading to   a negative 
disjoining pressure $\Pi_d$  with large amplitude. For sufficiently 
large  $n_0$ with small $H$, screening can also be achieved 
leading to a large positive  $\Pi_d$. 
 In this paper,  we are in  the nonlinear PB regime with addition of  NaOH  
only,  where the Debye length 
is longer than the Gouy-Chapman length \cite{Andelbook}.

Furthermore,  in a mixture solvent, 
  the dissociation on   
solid surfaces  can strongly 
depend on the local solvent composition $\phi$ as well as the solvation 
of ions in the bulk\cite{OnukiReview}. 
In our previous paper\cite{Oka2}, 
small  variations of  $\phi$ around the colloid surfaces 
induced  significant  changes in $\alpha$ and the ion distribution.
 Thus, there can be a strong coupling 
between $\phi$ and the ionization in mixture solvents, 
which  alters   adsorption and wetting  
on the  surfaces and the   interaction  among the 
colloidal particles.

In polyelectrolytes and charged gels 
\cite{Raphael,Oka0,Oka1,Netz,Muth,Bor,Olivera,Araki}, 
the dissociation on polymer chains is governed by 
the local environments and 
is highly fluctuating in  space and time. This  can strongly 
affect coil-globule transition and  phase separation, 
particularly when a second fluid component (cosolvent)  is added to a
water-like solvent. A first order 
phase transition of weak-to-strong ionization 
was also predicted on  ionizable rods in a mixture solvent\cite{Oka0}. 
In phase separation, $\alpha$ can be very different 
in the two phases\cite{Muth,Olivera,Oka1}.
With addition of   alcohol  to water, 
 precipitation of DNA has also 
been  observed \cite{Zimm}, where ionization of DNA 
is favored  in water-rich environments.

The organization of this paper is as follows. 
In Sec.II,  we will present   a  
 coarse-grained Ginzburg-Landau free energy functional, which 
includes the electrostatic contributions 
from the Stern layers and the dielectric film.
In Sec.III, we will present numerical results 
for silica-water interfaces for thick cells.   
  In Sec.IV, we will  
discuss the ionization on  carboxyl-bearing surfaces.  
In Sec.V, we will present results 
 for thin cells.

\section{Theoretical background}

\begin{figure}[tbp]
\includegraphics[scale=0.40]{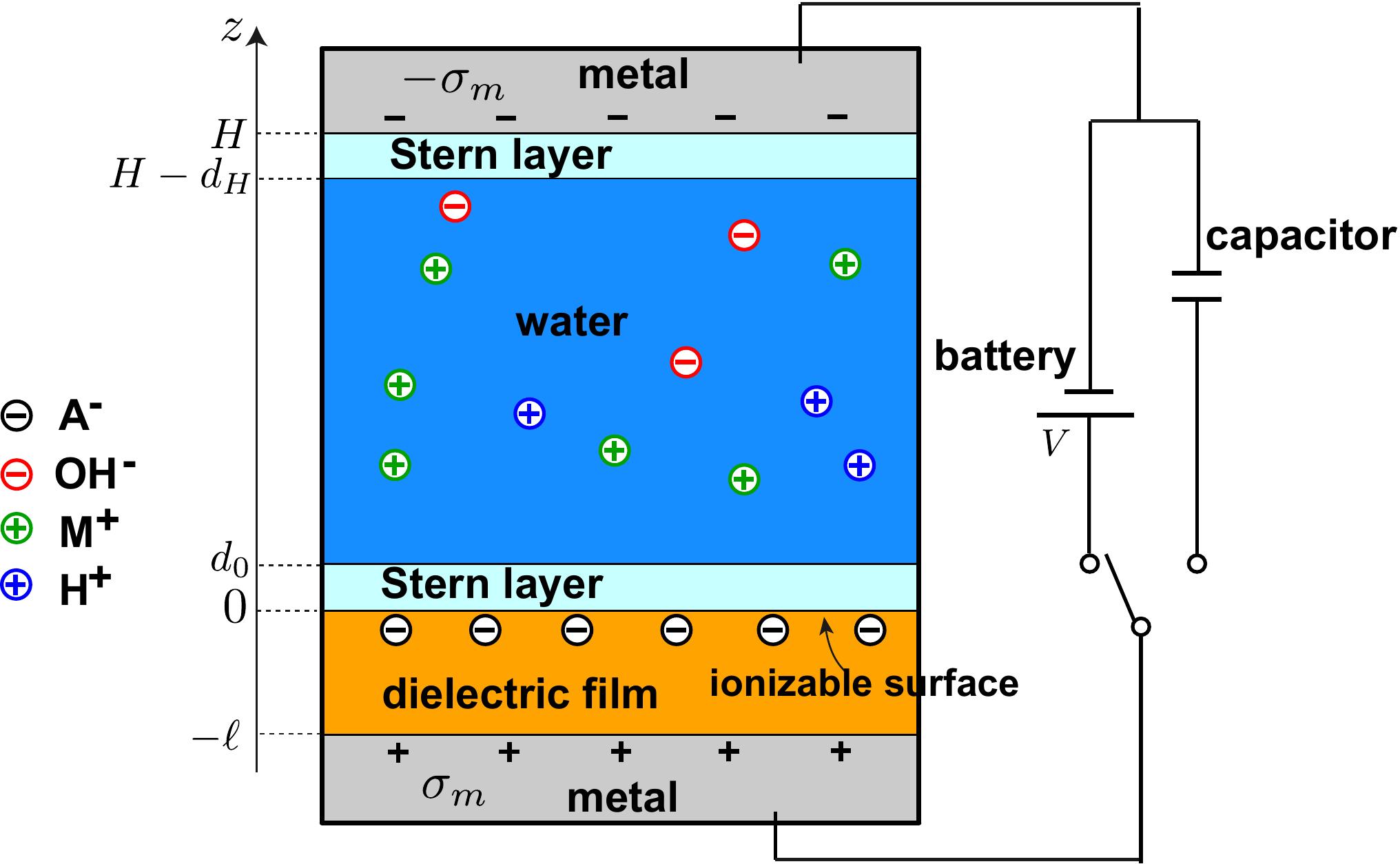}
\caption{Illustration  of geometry. 
Two parallel metallic plates 
are placed  at  bottom ($z< -\ell$) and top ($z>H$) 
with charge densities  $\pm \sigma_m$  and 
a potential difference  $V$. 
A dielectric film with thickness $\ell$ is 
 on  the bottom  metallic plate ($-\ell<z<0)$. 
At $z=0$, ionizable groups AH   with 
a  density $\Gamma_0$ dissociate into A$^-$ and H$^+$ with a fraction $\alpha$.
Water molecules are  under strong influence of the 
walls in   Stern layers ($0<z<d_0$ and $H-d_H<z<H$).
Protons  also come from dissociation  of water. 
MOH dissociates into M$^+$ and OH$^-$.    
If a battery is connected, 
$V$ can be controlled, where the metal surface 
chage density $\sigma_m$ fluctuates. 
If  it is disconnected, $\sigma_m$ becomes fixed. 
Furthermore, if a small capacitor is connected, 
$\sigma_m$ can be changed by a small fixed amount (see Appendix A).  
}
\end{figure}

We   illustrate our system  in  Fig.1, 
where a cell contains  liquid  water and ions  
in ambient conditions ($T=300$ K and $p=1$ atm)   
  in the region $0<z<H$ and 
 a  dielectric film is  in the region 
$-\ell<z<0$. For example, we suppose  a  ultra-thin  silica oxide film 
with    silanol  groups in  water.
To apply electric field to the system, 
we place  metallic   walls 
at the two ends in the regions $z< -\ell$ and $z>H$.  
In our theory, 
we can fix  the applied  potential difference $V$ 
 or the electrode  surface-change 
density $\sigma_m$.

The cell lengths in the lateral 
directions are much longer  than $H$ 
such that the edge effect is negligible. 
All the physical quantities are coarse-grained smooth variables 
depending  only on  $z$.   
Hereafter, the Boltzmann constant will be  set equal to 1.

\subsection{Chemical reactions in bulk and on surface}

In the cell, we  initially add a base   MOH  at a low density\cite{Yama}.   
It   dissociates into mobile univalent cations M$^+$  
and  hydroxide  anions ${\rm OH}^{-}$ as 
\be
{\rm MOH}\rightleftarrows {\rm M}^+ + {\rm OH}^- .
\en 
In addition, a very small fraction of water molecules  dissociate into 
 H$^+$ and OH$^-$ as \cite{Eigen}  
\be 
{\rm H}_2{\rm O} \rightleftarrows  {\rm H}^+ +{\rm OH}^-.  
\en 
We use the notation H$^+$, though  the protons 
 exist as hydronium ions H$_3$O$^+$ in liquid  water.  
The  local number  densities of MOH, M$^+$, OH$^-$, 
and H$^+$ are  written as $n_{\rm MOH}(z)$, 
$n_{\rm M}(z)$, $n_{\rm OH}(z)$, and $n_{\rm H}(z)$, respectively, which are 
coarse-grained smooth functions of $z$ in our theory.
 The charge density $\rho(z)$ is written as    
\be 
\rho= e(\nH+\nM -\nO) .
\en  
In chemical equilibrium, the mass action laws hold: 
\bea 
\nM \nO/\ns=K_{\rm b} ,\\
\nO\nH=K_{\rm w}, 
\ena 
where $K_{\rm b}$ and $K_{\rm w}$ are the dissociation 
constants. 
For    NaOH, we have    
$K_{\rm b}= 10^{-1}$mol$/$L$= 0.0 6$ nm$^{-3}$. 
However, $K_{\rm w}^{1/2} $ 
is much  smaller ($=10^{-7}$mol$/$L $=  6\times 10^{-8}/$nm$^{3}$). 

 In  a thick cell,   a homogeneous 
bulk region appears far from the walls,   
 where   $n_i$ assume  bulk values,  written as $n_i^0$.   
  They satisfy  Eqs.(4) and (5) and the charge neutrality condition 
$\nM^0+\nH^0=\nO^0$. We introduce the bulk density of 
 M atoms by 
\be 
n_0= \nM^0+\ns^0.
\en
See Fig.2(a) for  
$n_i^0$ vs $n_0$  for M$=$Na. 
For such a strong base  with large $K_{\rm b}$,  we 
can  well assume  $n_0 \ll K_{\rm b}$ to 
find 
\be 
\nM^0\cong n_0,\quad 
\ns^0 \cong \nO^0n_0 /{K_{\rm b}}
\ll n_0,  
\en 
where  $\nO^0\ll K_{\rm b}$. 
From   $n_0= (1+\nO^0/K_{\rm b})(\nO^0- \nH^0)$, 
we can  express $n_0$ in terms of $\nO^0$ as  
\be 
n_0\cong  \nO^0- \nH^0= \nO^0- K_{\rm w}/\nO^0,
\en  
In pure water, we have  $n_0=0$ and  $\nO^0=\nH^0=
K_{\rm w}^{1/2}$.  For   
$n_0\gg K_{\rm w}^{1/2}$,  we obtain 
 $\nO^0\cong  n_0$ and  $\nH^0\cong K_{\rm w}/n_0\ll K_{\rm w}^{1/2}$. 
In the following figures, 
 supposing a strong base,  we will  
use $n_0$ to represent the amount of the added base. 
However, our  theoretical results  
will be expressed in terms of $\nO^0$ and $\nH^0$ and 
will be  valid for arbitrary $K_{\rm b}$. In our figures 
 Eq.(8) holds, 
 so they can   be used even for 
 weak bases\cite{Yama1}  
 if $n_0$ is related to 
$\nO^0$  by Eq.(8) 

\begin{figure}
\includegraphics[scale=0.42]{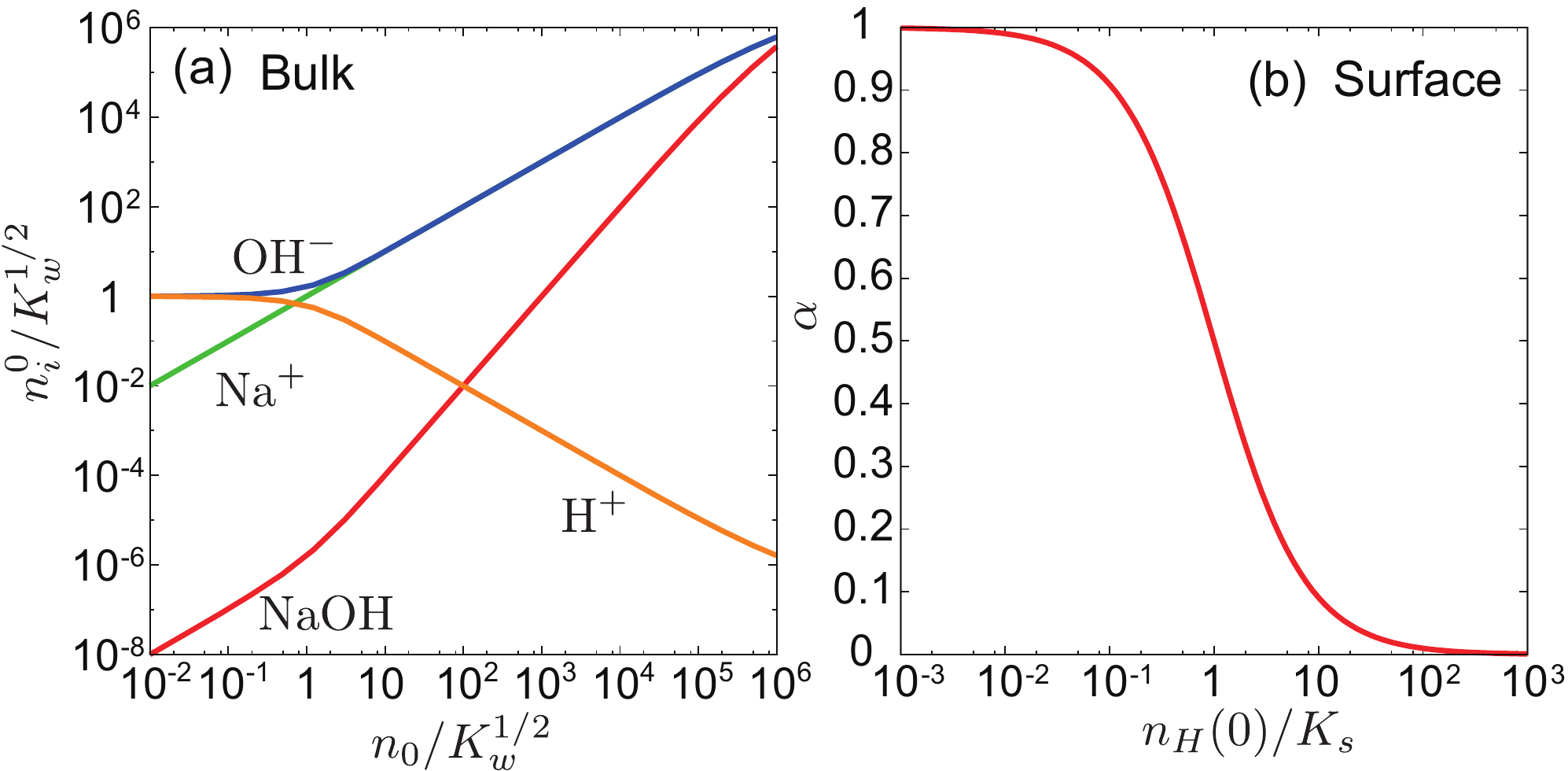}
\caption{ Chemical reactions in this paper. (a) Equilibrium bulk 
densities $n_i^0$ ($i={\rm NaOH}$, ${\rm Na}^+$, 
${\rm OH}^-$, and ${\rm H}^+$)  
divided by $K_{\rm w}^{1/2}$ 
vs $n_0/K_{\rm w}^{1/2}$, 
where  $n_0$ 
is the bulk density of Na atoms. (b) Equilibrium  $\alpha$ vs  $\nH(0)/K_s$ from Eq.(11). 
}
\end{figure}

On  the surface of the dielectric film at $z=0$, 
 ionizable groups  AH are distributed 
with a surface density $\Gamma_0$. 
No dissociation is assumed  on  the upper metal surface. 
Depending on the local pH near the surface, a fraction 
$\alpha$ $(0<\alpha<1)$ of these  groups  dissociate  as\cite{Par}   
\be
{\rm AH} \rightleftarrows {\rm A}^- +{\rm H}^+ .  
\en
Here,  A$^-$ anions remain  on the 
surface $z=0$, while the dissociated protons are mobile 
in water.  
The  surface charge density due to  A$^-$  at $z=0$ is  
\be 
\sigma_{\rm A}= -e\Gamma_{0}  \alpha.
\en 
Let $\nH(0)$ be  the proton density 
immediately close to the dielectric film in water. Then, 
the surface mass action law in chemical equilibrium 
is expressed as \cite{Par}    
\be 
 \nH(0) \alpha  /(1-\alpha)= K_s, 
\en 
where   $K_s$ is the  surface  dissociation constant.  
See Fig.2b  for $\alpha$  vs $\nH(0)/K_s$.  
Since $\alpha=1/[1+\nH(0)/K_s ]$, 
$\alpha$  tends to 1 (to 0) 
if  $\nH(0)$ is much smaller (larger) than $ K_s$.

We define the  particle  numbers $N_i= \int_0^H dz n_i(z)$  
 (i= MOH, M, OH, H)   per unit area 
 in the cell.  Since  the hydroxide and proton numbers due to 
the autoionization coincide, we have 
\be 
N_{\rm OH}-N_{\rm M}=  N_{\rm H}- \Gamma_{0}  \alpha. 
\en 
It follows  the overall  charge neutrality condition, 
\be 
 \int_0^H   dz  \rho(z)+\sigma_{\rm A}=0.
\en  
In our theory, 
 a reservoir can be attached or 
 the  system can be closed.  
In the latter case, 
the total number of the M atoms is fixed as 
\be
 N_{\rm M}+N_{\rm MOH}=H {\bar n},
\en  
where $ {\bar n}$  is   the initial density  
  of  the  added base.

In Sec.III, we consider  silica oxide  surfaces 
with  silanol groups AH= SiOH and 
 set \cite{BehR,Beh0,Beh,Wes,Hie,Fuer,Leckie} 
\bea 
K_s&=& 10^{-7.3}{\rm mol}/{\rm L} 
= 3.0 \times 10^{-8}/{\rm  nm}^{3}   ,\nonumber\\
\Gamma_0&= & 8.0/{\rm nm}^2 \quad ({\rm silica}~{\rm surface}),
\ena 
where    $K_s$  is 
very small $(=0.5K_{\rm w}^{1/2})$. The pH and  pK values are 
 defined by $\nH^0 =10^{-{\rm pH}}$ mol$/$L (in the 
bulk region) and 
 $K_s= 10^{-{\rm pK}}$ mol$/$L. Then, pK$=7.3$,  
 $\nH^0/K_s =10^{7.3-{\rm pH}}$, and   $\nO^0/K_s =10^{{\rm pH}-6.7}$  here. 
The value of $\Gamma_0$ in Eq.(15) 
is obtained for nonporous, fully hydrated silica, 
so it   is large. Experimental  values  
of $\Gamma_0$   strongly depend  on surface preparation\cite{Beh}.
 It is also known that  the adsorption 
 SiOH$+$H$^+$ $\rightleftarrows$ SiOH$_2^+$ 
 takes place in  high acidity (low pH) conditions 
\cite{Fuer,Hie}, leading to a zero-charge surface state at a pH about 2. 
See Sec.IV for analysis on ionization on 
carboxyl-bearing surfaces.

\subsection{Electric potential and Stern layers}

 As the electrostatic boundary condition, 
we may fix   the  potential difference between the electrodes: 
\be
V= \Phi(0)-\Phi(H)+  \sigma_m  /C_d.
\en  
We may also fix the surface charge density $\sigma_m$ 
on the lower metal surface at $z=-\ell$ (see Appendix A). 
In the dielectric film, the electric field is given by 
$4\pi  \sigma_m/\ve_d$, where  $\ve_d$  is the film 
dielectric constant. 
This yields the potential change  $\Phi(-\ell)-\Phi(0)
= \sigma_m  /C_d$, 
where the film  capacitance per unit area is  written as 
\be 
C_d= \ve_d/4\pi \ell.
\en 
In the cell outside  microscopic Stern layers (see below), 
$\Phi(z)$  obeys the Poisson equation, 
\be 
-\ve_0 d^2 \Phi/dz^2  = 4\pi \rho, 
\en 
where  $\ve_0$ 
is the solvent dielectric constant. 
The electric field is given by $E= -d\Phi/dz$. 

Generally,    the electric 
potential can  change  noticeably  
  across  a microscopic  Stern layer 
at  a solid-water interface\cite{Is,Butt,Hunter,Kolb}.
We mention molecular dynamics 
simulations  on this effect \cite{Madden,Hautman,Takae,Yeh,Hender}. 
In our case, there are two such layers 
at the bottom and top. 
For simplicity, we  assume no specific ion adsorption. Then, 
 the ion amounts   in the layers 
are  negligibly small for small  bulk ion densities
\cite{Beh0,Beh1,Beh,Hender}.  
Denoting   the  layer thickness   as 
$d_0$ at $z=0$ and as $d_H$ at $z=H$, we  assume    
 linear  relations, 
\bea  
V^{\rm S}_0 &=& \Phi (0)  -\Phi (d_0) =
 (\sigma_m+ \sigma_{\rm A})/C_{0},\nonumber\\ 
 V^{\rm S}_H &=& \Phi (H-d_H)  -\Phi (H) 
= \sigma_m /C_H. 
\ena 
For simplicity, the surface capacitances   $C_{0}$ and $C_H$ are 
taken to be those  
in the limit of small ion densities. 
Here, we can define the polarization $P(z)$ microscopically 
such that the electric induction 
$D(z)= E(z)+4\pi P(z)$ is continuous through the 
layers\cite{Takae}, where   $E(z)$ and $P(z)$ change  
abruptly in the layers\cite{comment1}.  We write  $D(z)$ 
at the bottom and the top as   
\bea 
&&D(d_0)= \ve_0 E(d_0) = {4\pi}   (\sigma_A+\sigma_m).
\nonumber\\ 
&&D(H-d_H)= \ve_0
E(H-d_H) = {4\pi} \sigma_m,
\ena
These constitute the boundary conditions of 
the Poisson equation (18). The effective surface charge density 
is $\sigma_{\rm eff}= \sigma_{\rm A}+\sigma_m$ at the bottom.  
Furthermore, even without applied field ($\sigma_m=0$), 
the previous simulations \cite{Takae,Madden,Hautman,Yeh,Hender} 
have  shown the presence of  small  potential drops    
  at solid-water interfaces due to  
  the anisotropy of  water molecules . 
We  neglect this intrinsic,  surface effect in this paper. 

The total potential difference $V$ is now written as 
\be 
V= [\Phi(d_0)-\Phi(H-d_H)]+ \frac{\sigma_{\rm A}+\sigma_m}{C_0}
+\frac{\sigma_m}{C'},
\en  
where the first  term arises from  the mobile 
ions and  
\be 
C'= ( 1/C_H+1/C_d)^{-1} .
\en  
If we neglect the image interaction\cite{Landau}, 
the electrostatic free energy $F_e$ appropriate at fixed $\sigma_m$ 
is  the space integral of $DE/8\pi$ in the whole region ($-\ell<z<H)$. Then,  
\bea
\hspace{-2mm} 
F_e&=&\int_{d_0}^{H-d_H}\hspace{-2mm}  dz \frac{\ve_0 E^2}{8\pi}
+\frac{(\sigma_A+\sigma_m)^2}{2C_0}
 +  \frac{\sigma_m^2}{2C'} \nonumber\\
&=&\frac{1}{2}\int_{0}^{H}\hspace{-2mm}  dz \rho \Phi+ \frac{1}{2} 
\sigma_{\rm A} \Phi(0) + \frac{1}{2} \sigma_mV .
\ena 
We rewrite  the integral in the first line as 
$[\Phi(d_0)D(d_0)-\Phi(H-d_0)D(H-d_0)]/8\pi 
 + \int dz \Phi\rho/2$ 
and use Eq.(20). 
It then follows  the second line, 
where  the integration region $d_0<z< H-d_H$ has 
been changed to $0<z<H$  and 
$\sigma_{\rm A}$   is related to $\rho$ by Eq.(13).
 Note that 
the integrals of $\rho$ 
in   the Stern layers are assumed to be negligible.

We also consider small  changes $\rho \to \rho+\delta\rho$ and  
$\sigma_m \to \sigma_m+\delta\sigma_m$. 
The incremental change in  $F_e$  is written as  
\be 
\delta F_e=  \int_0^H dz \Phi\delta\rho + \Phi(0) \delta\sigma_A  
 + V\delta\sigma_m ,
\en  
where $\delta\sigma_A=- \int_0^Hdz \delta\rho(z)$  from Eq.(13) 
and the last  term vanishes at fixed $\sigma_m$. 
On the other hand, at fixed $V$, 
the appropriate free energy 
is $F_e  -\sigma_m V$, whose differential form is given by 
the first two terms in Eq.(24). Here, we neglect the ion-ion 
correlation due to the  fluctuations in the 
$xy$ plane and the image interaction 
between the ions and the image charges in the solid regions. 

In our  analysis,  we  use the following capacitance values. 
(i) For our dielectric film, 
we assume  $\ve_d=4$ and $\ell=1.05$ nm 
to obtain  $C_d=0.0335$ F$/$m$^2$ from Eq.(17). On the other hand, 
 in analysis of electrowetting,  Klarman and Andelman \cite{Andel1} 
assumed a much smaller value, 
$C_d=4.4\times 10^{-6}$ F$/$m$^2$,  
for a film with $\ve_d=2.67$ and $\ell=5$ $\mu$m. In their  case, 
the relation  $V\cong \sigma_m/C_d$  held  nicely for not 
very small $V$.  
(ii) Supposing  a silica-water interface \cite{Hie,Beh0,Beh1,Beh}, 
we   set   
 $C_0=2.9$ F$/$m$^2$.   
(iii)  The metal-water  
capacitance\cite{Kolb,Madden} 
has been  observed   in the range   
$0.2-0.5$ F$/$m$^2$, so we set   $C_H=0.3$ F$/$m$^2$. 
Then,  $1/C_H = 0.11/C_d$ in our case. 
Via  microscopic simulations 
 \cite{Madden,Hautman,Takae},   
the surface capacitance for a metal-water interface 
has  been calculated   in a range  of  $0.05-0.1$F$/$m$^2$ 
with  a layer thickness about 5$\rm\AA$ (where 
water molecules are depleted).

\subsection{Free energy functional}
We   set up  the  Helmholtz free energy functional $F$  
for a  cell with    $d_0 \ll H$ and 
$d_H\ll H$ in the mean field theory. 
At fixed $\sigma_m$, it consists of three parts  as   
\be 
F=   F_e   +F_b+ F_s. 
\en 
At fixed $V$, we should replace $F_e$ 
by $F_e -\sigma_m V$. 
Here,   $F_b$ is the contribution from the 
solute particles and $F_s$ 
is that for the surface ionization: 
\bea
\hspace{-6mm} 
{F_b}/T &=& \int_0^H  \hspace{-1mm}
dz \bigg[ \sum _{i} n_i[  \ln (n_i \lambda_i^3) -1]+\Delta_{\rm b}\nM 
\nonumber\\
&& \hspace{5mm}   +\Delta_{\rm w}(\nO-\nM) \bigg],\\
\hspace{-2mm}
F_s/T \Gamma_0  &=& 
\alpha \ln \alpha +(1-\alpha )\ln (1-\alpha )+ \Delta _\sigma \alpha    ,
\ena 
where  $\lambda_i (\propto T^{1/2}$) 
 is the thermal de Broglie length  of the particle species $i$. The 
$T\Delta_{\rm b}$, $T\Delta_{\rm w}$, and 
$T\Delta_s$ are the dissociation free energies  
for the chemical reactions in Eqs.(1), (2), and (9), respectively, 
which   arise  from the microscopic  
interactions. 
The above form of $F_s$ has long been used in the literature
\cite{Raphael,Muth,Bor,Oka1,Oka2,Ohshima}, where 
the first two terms are  entropic contributions.

We define the chemical    potentials of the solute 
particles by $\mu_i= \delta F/\delta n_i$ at fixed  $\alpha$. 
Then,  Eqs.(25)-(27) give  
\bea 
&&\hspace{-3mm} \mu_{\rm MOH}/T 
= \ln [\ns\lsa^3], \\
&&\hspace{-3mm}\mu_{\rm M}/T
= \ln [\nM \lM^3] +\Delta_{\rm b}-\Delta_{\rm w}+U, \\
&&\hspace{-3mm}\mu_{\rm OH}/T=\ln [\nO\lO^3]+\Delta_{\rm w}-U ,\\ 
&&\hspace{-3mm}\mu_{\rm H}/T= \ln [\nH\lH^3]+U, 
\ena
where  $U(z)=e\Phi(z)/T$ is  the normalized potential.

In  equilibrium, we  require  the charge neutrality 
condition (12).  Here, we can either assume 
  Eq.(14)  with a constant $\bar n$ in a closed cell 
or  attach a reservoir 
to the cell with a common $\ms$ 
(see Appendix B)\cite{Par}.  
In these  cases,   we should minimize  the grand potential,  
\bea 
\Omega &=&F -h_0( N_{\rm M}+N_{\rm MOH})\nonumber\\
&-& h_1( N_{\rm H}-N_{\rm OH}+N_{\rm M}-\Gamma_0\alpha),
\ena 
where    $h_0$ and $h_1$ are  homogeneous constants. 
Requiring    $\delta \Omega/\delta n_i=0$, we obtain 
 the  chemical equilibrium conditions for the mobile particles: 
\bea
&&\ms =\mM+\mO= h_0, \\
&&\mH =-\mO = h_1.
\ena 
With these relations, $\Omega$ can  be expressed  as  
\be 
\Omega = F- \sum_{i\neq {\rm H}}\mu_i 
 N_i- \mu_{\rm H}(N_{\rm H}- \Gamma_0\alpha). 
\en 
See Appendix D for more detailed expressions of $\Omega$.

Assuming  the equilibrium conditions in Eqs.(33) and (34)  for $n_i(z)$, 
we may  treat  $\Omega=\Omega(\alpha)$  as  a function of $\alpha$. 
Then, its minimization with respect to  $\alpha$ gives the equilibrium $\alpha$. From Eq.(27) we  obtain  its derivative, 
\be 
\frac{1}{T\Gamma_0} 
\frac{d\Omega}{d\alpha} = 
\ln(\frac{\alpha}{1-\alpha}) +\Delta_s +  \ln [\nH(0)\lH^3] .  
\en 
Using  Eqs.(33) and (34)  and setting 
$d\Omega/d\alpha=0$,  we derive 
the  chemical equilibrium conditions (4), (5), and (11) with 
the dissociation constants, 
\bea 
&&K_{\rm b}= (\lsa/\lH \lO)^{3} \exp({-\Delta_{\rm b}}),\\
&&K_{\rm w}= (\lH\lO)^{-3} \exp({-\Delta_{\rm w}}),\\
&&K_s=  \lH^{-3}\exp({-\Delta_s}).
\ena 
With Eq.(39) the right hand side of Eq.(36) becomes 
$\ln[\alpha \nH(0)/(1-\alpha)K_s]$, leading to Eq.(11) 
in equilibrium. 

If we use $K_b$ for NaOH and $K_s$ for  silica, 
 Eqs.(37)-(39) 
give   $\Delta_{\rm b}=8.4$, $\Delta_{\rm w}=51$, 
and $\Delta_s=24$.  
These large sizes of $\Delta_{\rm w}$ and  $\Delta_s$ 
indicate that the autoionization in water 
and the dissociation on a silica oxide  surface are 
rare activation processes. 
On the other hand,    $\Delta_{\rm b}$ is relatively small 
 such  that  NaOH mostly dissociates in water.

We previously presented  free energies  
with  variable charges in mixture solvents  
for  colloidal  particles\cite{Oka2},  rods\cite{Oka0}, 
and polyelectrolytes \cite{Oka1} (without applied   field). 
In such systems, if   $\Delta_s$ depends on the composition,  
ionization and wetting transitions   
are coupled. 

\subsection{Space-dependence of ion densities}
From Eqs.(28)-(31),  $\ns$ is a  homogeneous constant. 
The ion  densities can be  expressed in terms of $U(z)$ as   
\bea 
&& \nO(z)= n^0_{\rm OH} e^{U(z)}, \quad 
 \nH(z)=n^0_{\rm H} e^{-U(z)},\nonumber\\
&&\nM(z)=n^0_{\rm M}e^{-U(z)}.
\ena 
where $n_i^0$ are constants 
with $\nM^0+n^0_{\rm H}=\nO^0$. See Appendix B 
for more details on these expressions. 
From Eq.(18)   we obtain the  PB equation, 
\be 
d^2 U/dz^2 = \kappa^2  \sinh (U), 
\en  
We define the Debye wave number  $\kappa$  by   
\be 
\kappa= (8\pi \ell_B \nO^0 )^{1/2}
=10^{-3}({\nO^0/K_{\rm w}^{1/2}})^{1/2}  /{\rm nm}.
\en 
where  $\ell_B=e^2/\ve_0 T= 7.0~{\rm \AA}$ is the Bjerrum length. 
The inverse $\kappa^{-1}$ is the Debye length, 
which  is  long  here, but it  can be shortened 
 if we add  a  salt such as KCl.  
In such  cases, $\nO^0$ in Eq.(42) should be replaced by 
the total anion density $n_{\rm b}^0$ in the bulk.

 For $\kappa H\gg 1$,  
 $n^0_i$ are the bulk ion densities 
far from the walls as in Eqs.(6)-(8), where 
the charge neutrality condition holds.  
However,  even for not large   $\kappa H $, we can use 
Eq.(40) with  well-defined $n_i^0$. 
Under Eq.(14) in a closed cell, $\nM^0 $ is 
determined by   
$\nM^0 \int_0^H dz e^{-U}/H +n_{\rm MOH}={\bar n}$ (see Eq.(58)). 
If we attach a reservoir to the cell with a common $\ns$, 
$n_i^0$ in Eq.(40)  are simply equal to  the ion densities in the reservoir 
(see Appendix B).

 The protons are assumed to 
penetrate  into  the Stern layers.   
From Eq.(19) the ratio of its values at $z=0$ and $d_0$ 
is given by  the Boltzmann factor $\exp(-eV_0^{\rm S}/T)$, so  
\be
\nH(0)/\nH(d_0) = 
 \exp[-{e( \sigma_{\rm A}+\sigma_m)}/{C_0T}].
\en 
This relation was proposed    by Behrens {\it et al.}\cite{Beh0,Beh1,Beh}, 
but it should  be checked with  microscopic simulations.



\subsection{Osmotic and disjoining pressures}
In our system, the osmotic   pressure  
$\Pi$ is  defined in the bulk 
region $d_0<z<H-d_H$  and is given by\cite{Andel3} 
\be 
\Pi=T \sum_i n_i(z) - \frac{\ve_0}{8\pi }  E(z)^2, 
\en  
where the first term is the partial pressure 
of the  solute particles 
and the second term is the $zz$ component of 
the Maxwell stress tensor\cite{Landau}. In equilibrium, 
$\Pi$   is a constant independent of $z$. In fact, 
$d\Pi/dz=0  $ from Eqs.(18) and (40).  Here, we 
assume   $H> d_0+d_H\sim 1$ nm. 

Let us treat  the equilibrium (minimum)  
value of $\Omega$  under Eqs.(11), (33), and (34) as a function of 
 the cell width $H$  in the fixed charge 
or the fixed potential condition. 
In Appendix C, we will derive the following   relation, 
\be 
\Pi=- \frac{\p}{\p H}\Omega  . 
\en  
A similar  formula holds for the force between 
two colloidal particles with  $H$ being  their separation 
distance\cite{Oka3}. 
As will be shown 
in Appendix C,  Eq.(45)  holds when 
$\ns$ or   $h_0=\mu_{\rm MOH}$ in Eq.(33) is fixed in the derivative. 
This   is equivalent to fixing  
 $\nM\nO$ from Eq.(4). It  can be realized if we attach a reservoir 
to the cell.  On the other hand, if the cell is closed and 
$\bar n$ in Eq.(14) is fixed, 
Eq.(C5) in Appendix C leads to  
 $\Pi=-\p F/\p H$.

We    suppose   a  reservoir without applied  field.  The 
  $\ns$ in the reservoir   is the same as that 
in the cell.     Then, 
the disjoining pressure is the difference 
 $ \Pi_d= \Pi- \Pi_r^0$, where  $\Pi_r^0$ is 
 the  osmotic pressure in the reservoir (see  Appendix B). 
   Using $ n^0_{\rm OH} $  in Eq.(40),  we find\cite{Andel3,AndelEPL,Is}   
\be 
\Pi_d= T n^0_{\rm OH} [2\cosh(U)-2-  \kappa^{-2}|d{U}/dz|^2].   
\en 
Thus, $\Pi_d= -\p (\Omega +H\Pi_r^0)/\p H$ 
 at fixed $\ns$. In  the presence 
of other kinds of monovalent ions, 
 $\nO^0$ in Eq.(46) should be replaced by 
the total anion (cation) density $n_{\rm b}^0$ 
in the reservoir, as stated below Eq.(42). 

\begin{figure}[tbp]
\includegraphics[scale=0.4]{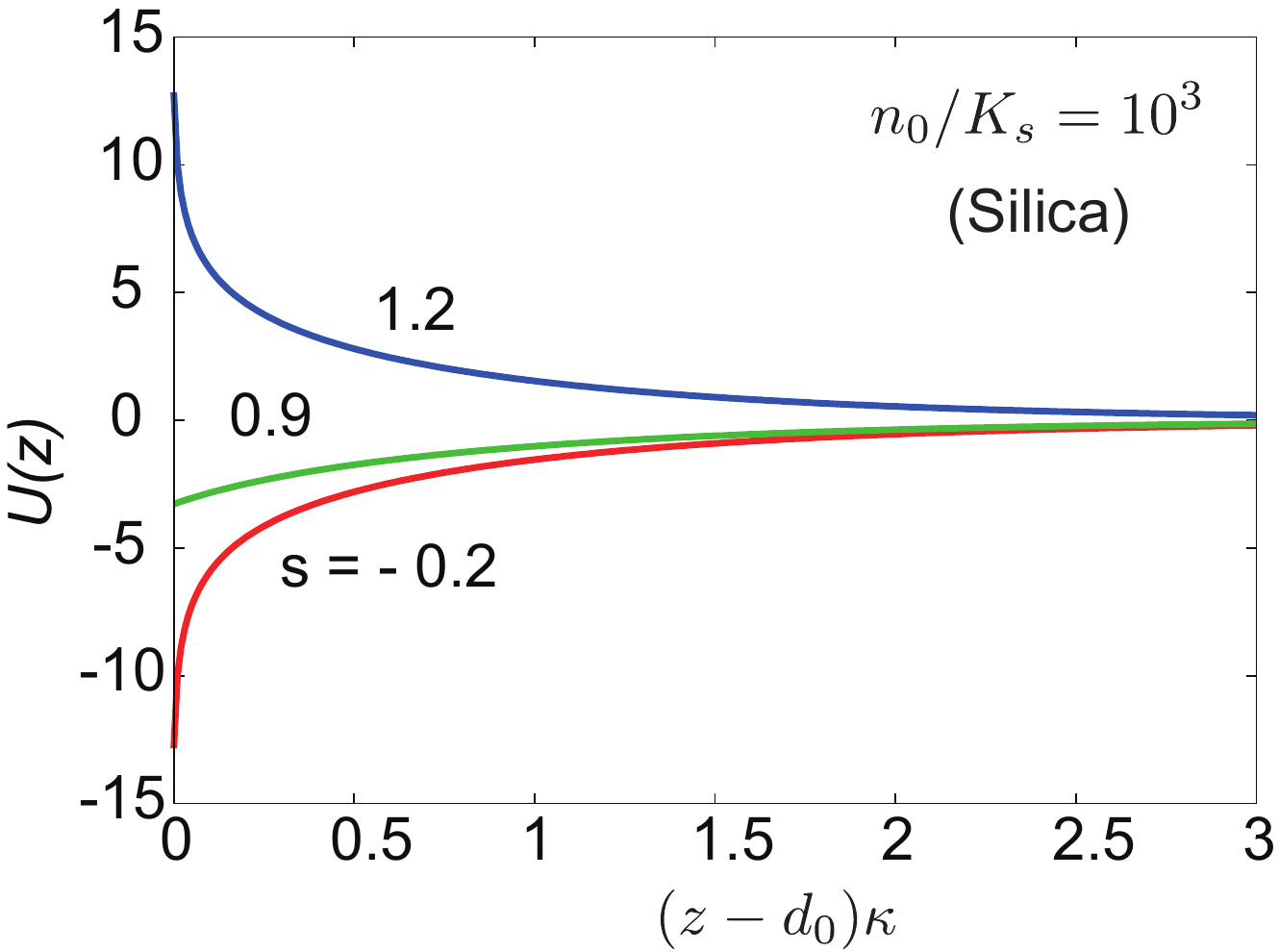}
\caption{ Normalized electric 
 potential $U(z)$  above the Stern layer 
$(z>d_0)$ near the bottom wall 
  at $A_1=1550$ ($n_0/K_s= 10^3$ for silica), 
where $\alpha$ depends on  $s=\sigma_m/e\Gamma_0$. 
For three curves the initial slope at $z=d_0$ 
is $2A_1 u$ with  $u=\alpha-s$, 
where  $(s,\alpha)= (-0.20, 2.2\times 10^{-5})$, 
$(0.90, 0.902)$, and $(1.2, 1.0)$ from below. 
Thus,  $u= 0.20$, $0.002$, and $-0.2$, respectively.
For the middle curve ($s=0.90$), 
$0<u\ll 1$  due to self-regulation.
For the other curves the initial slope is steep.}
\end{figure}

\section{Ionization on silica surface at large wall separation  } 

We consider equilibrium 
in a thick cell with $H\gg \kappa^{-1}$, $d_0$, and $d_H$.    
In the middle region $z\gg  \kappa^{-1}$ and  $H-z\gg  \kappa^{-1}$, 
the electrolyte is homogeneous with 
bulk   ion densities $n_i^0$.  

\subsection{Basic solution for large $H$}

For $ H \gg \kappa^{-1}$,  
we impose the semi-infinite boundary condition:  
$U(z) \to 0$ far from the walls. Near the bottom and the top, 
the profiles of the diffuse layers are given by   
\bea 
&&\hspace{-15mm}
U(z) =  2 \ln \bigg 
[\frac{1-\eta e^{-\kappa (z-d_0)}}{1+\eta e^{-\kappa( z-d_0)}}
\bigg] \quad\quad ({\rm bottom})
 \nonumber\\
&&\hspace{-5mm}
= 2 \ln \bigg [\frac{1-\zeta 
 e^{\kappa (z-H+d_H)}}{1+\zeta e^{\kappa (z-H+d_H)}}\bigg] 
\quad ({\rm top}). 
\ena 
From Eq.(20) the coefficients   $\eta$  and $\zeta$ are determined by 
\bea
&&2\eta/(1-\eta^2)= -(2\pi \ell_B/e\kappa )
(\sigma_{\rm A} + \sigma_m),\\
 && 2\zeta /(1-\zeta^2)= (2\pi \ell_B/e\kappa )\sigma_m.  
\ena
We define dimensionless parameters $A_1$, $A_2$, 
$u$, and $s$ by  
 \bea 
&&\hspace{-5mm} 
A_1=2\pi \ell_B\Gamma_0/\kappa=\Gamma_0 (\pi\ell_B/2\nO^0)^{1/2}
 ,\\
&&\hspace{-5mm} 
 A_2= e^2 \Gamma_0/C_0T, \\
&&\hspace{-5mm}
u= -(\sigma_{\rm A} + \sigma_m)/e\Gamma_0= \alpha- s ,\\
&&\hspace{-5mm}
s=  \sigma_m/e\Gamma_0.
 \ena 
Then, the  right hand side  of Eq.(48) is   $A_1u$ 
and  that of Eq.(49) is $A_1s$. 
 From Eq.(20),  $dU/dz$ is 
$2A_1u $ at $z=d_0$ and is 
$-2A_1s$ at $z=H-d_H$ as the boundary conditions. 

In terms of these parameters   the  total  potential difference 
in Eq.(21) is rewritten as   
\be
\frac{e}{T}V = - g(A_1u)  
- A_2 u+ g(A_1s)   
+  \frac{e\sigma_m}{TC'} .
\en  
where $g(x)=  2\ln[ \sqrt{1+x^2}+x]$. Then, $g(-x)=-g(x)$. 
In Eq.(54) the first  term  ($=U(d_0)$) 
arises from the lower diffuse layer, 
the second  from the lower Stern layer, 
the third  ($= -U(H-d_H)$) from the upper diffuse layer,  
and the  last term is due to the dielectric film and the upper Stern layer 
with Eq.(22).
Here, the  Debye-H$\ddot{\rm u}$ckel  theory 
is valid  only for  $|u|\ll A_1^{-1}$ at the bottom 
 and  for $|s|\ll A_1^{-1}$ at the top. 
For silica, we have  $A_1 = 4.9\times 10^4  
(K_s /\nO^0)^{1/2}$ and $A_2= 17$. 
Thus,  $A_1\gg 1$ for realistic  $\nO^0$ because of large $\Gamma_0$ 
and small $\kappa$.

Ninham and Parsegian \cite{Par} set 
    pK=4.8,   $\Gamma_0=1$ and $0.25 /$nm$^2$, 
 and  $\kappa=1.29/$nm with a salt added.
As a result, $A_1$ in Eq.(50) was of order 1 
and $\kappa^{-1}$  did not exceed 
the  Gouy-Chapman length  
(see Eq.(56) below).


\subsection{Profiles in the nonlinear PB regime}

There appear three cases with increasing $s$: (i) 
 $s<0$ and $u>0$  with $dU/dz>0$, 
(ii)  $s>0$ and $u>0$  
with $U(z) $ lower near the walls than in the middle, 
  and  (iii) $s>0$ and $u<0$  with 
$dU/dz<0$. We are interested in  the nonlinear   
behavior of $\alpha$ in case (ii), 
since $\alpha\to  0$ in  case (i) 
and  $\alpha\to  1$ in  case (iii).

In Fig.3, we display $U(z)$ near the bottom 
for $A_1=1.55\times 10^3$ (at $n_0/K_s= 10^3$ 
for silica), where $\alpha$ is determined 
for given $n_0$ and $s$. Then,   $(s,\alpha)=(-0.20, 2.2\times 10^{-5})$, 
$(0.90, 0.902)$, and $(1.2, 1.0)$.   
In the case  of $s=0.9$,   $u$  is small ($=0.002$) 
or $(\sigma_{\rm A}+\sigma_m)/\sigma_m\cong -0.002$, 
but $A_1 u=3.1$  in the nonlinear PB regime. 
 We shall see that this charge cancellation 
 is a universal effect. 
     
Let us consider the profile of $U$ and $\nH$ at 
the bottom in the nonlinear PB regime $A_1 u \gg 1$ with $u>0$. 
The coefficient  $\eta$ in Eq.(48) is close to 1 as 
$\eta\cong 1+ 1/A_1 u$, so 
\bea 
&&U(z) \cong 2\ln [(\kappa (z-d_0 + \ell_{\rm GC})/2], \nonumber\\
&&\nH(z)/\nH^0 \cong  4 \kappa^{-2} 
 (z-d_0 +  \ell_{\rm GC})^{-2} .
\ena 
Here, $\ell_{\rm GC}$ is 
  the Gouy-Chapman length\cite{Andelbook} at the bottom,  
\be 
\ell_{\rm GC}= (\kappa A_1| u|)^{-1}
= e/(2\pi\ell_B |\sigma_{\rm A}+\sigma_m|),
\en 
where we assume  $\ell_{\rm GC}\ll \kappa^{-1}$. 
The profiles change   on the scale of $\ell_{\rm GC}$.  
 In this case, $\nH(d_0)$ is larger than $\nH^0$  as 
\be 
\nH(d_0)/\nH^0 \cong (2A_i u)^2 .
\en  
The ratio $\nM (d_0)/\nM^0$ is also 
given by the right hand side of Eq.(57), which is  
important for $\nM^0\cong \nO^0\gg \nH^0$.

\begin{figure}[tbp]
\includegraphics[scale=0.455]{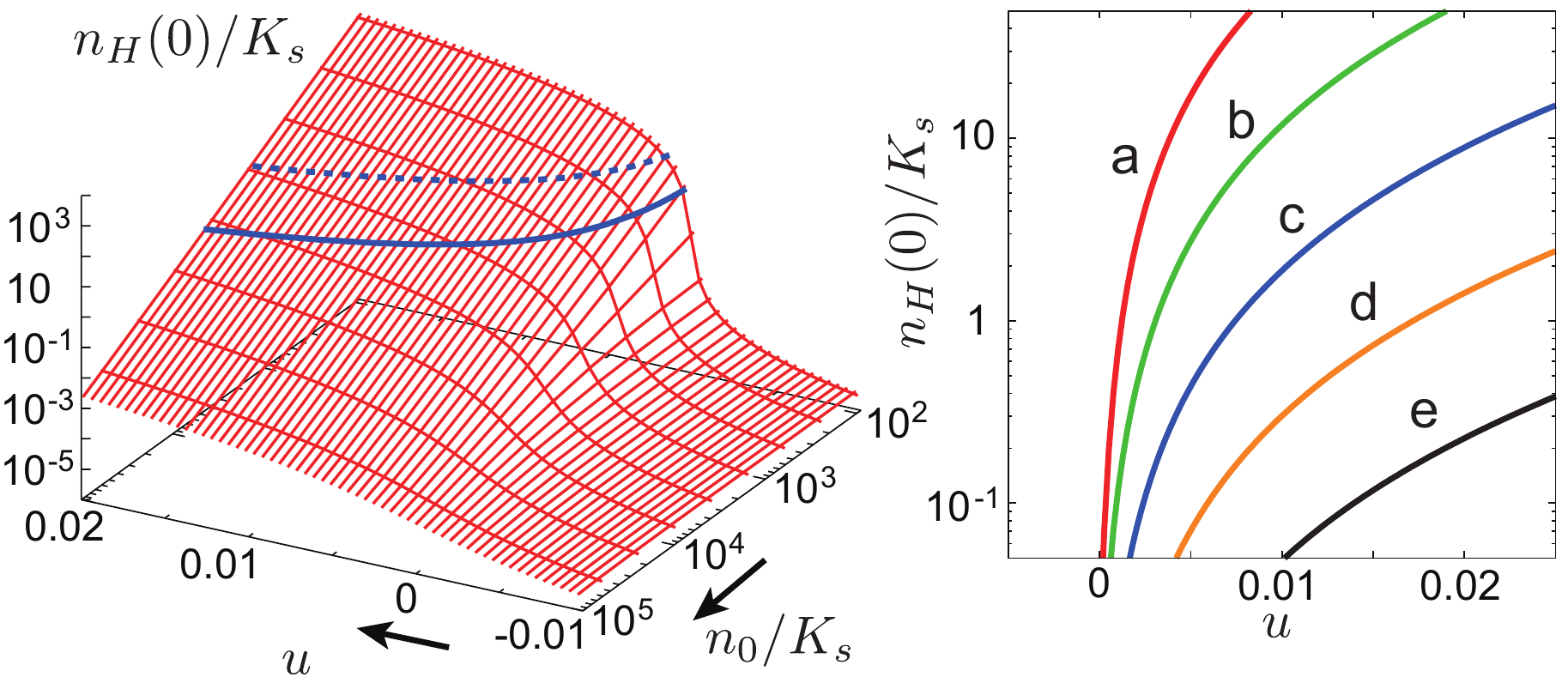}
\caption{ Left: Ratio ${\nH(0)}/K_s $ from  Eq.(59)
in the $u$-$n_0/K_s$ plane  with  
$u$ in Eq.(52), which is 
$(1-\alpha)/\alpha$ in equilibrium. On the surface,  
it is $1$ on the lower bold line and  is 10 
on the upper dotted line, where $\alpha$ is 0.5 and 0.09, respectively .  
 Right: ${\nH(0)}/K_s $ vs $u$ with 
 $n_0/K_s$ being (a) $2.5\times 10^2$, 
(b) $6.3\times 10^2$, (c) $1.6\times 10^3$ (right), 
(d) $4.0\times 10^3$, and (e) $1.0\times 10^4$ from above, 
where increase is steep for $n_0/K_s < 10^{-4}$.
 }
\end{figure}

\begin{figure}[tbp]
\includegraphics[scale=0.4]{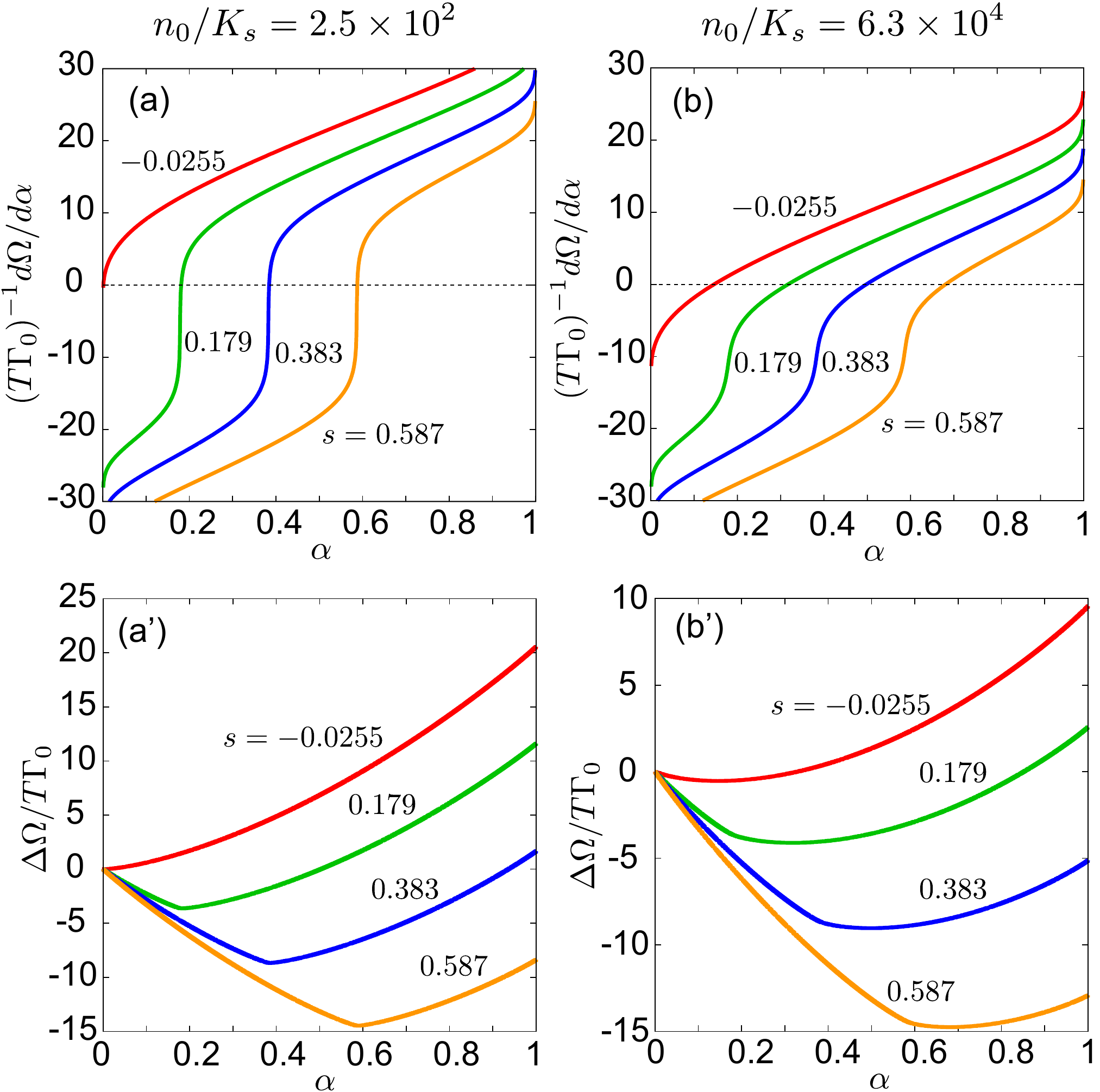}
\caption{ Left: $d \Omega (\alpha)/d\alpha$ and 
$\Delta\Omega= \Omega(\alpha) -\Omega(0)$ 
divided by $T\Gamma_0$ as functions of  $\alpha$ 
for $s=\sigma_m/e\Gamma_0=  -0.0255, 0.179, 0.383$, and $0.587$ 
  at  $n_0/K_s= 2.5\times 10^2$. 
Minimization  of  $\Omega(\alpha)$ gives 
equilibrium $\alpha$.  
 Right: those   at $n_0/K_s=6.3\times 10^4$.  
  }
\end{figure}

\begin{figure}[tbp]
\includegraphics[scale=0.41]{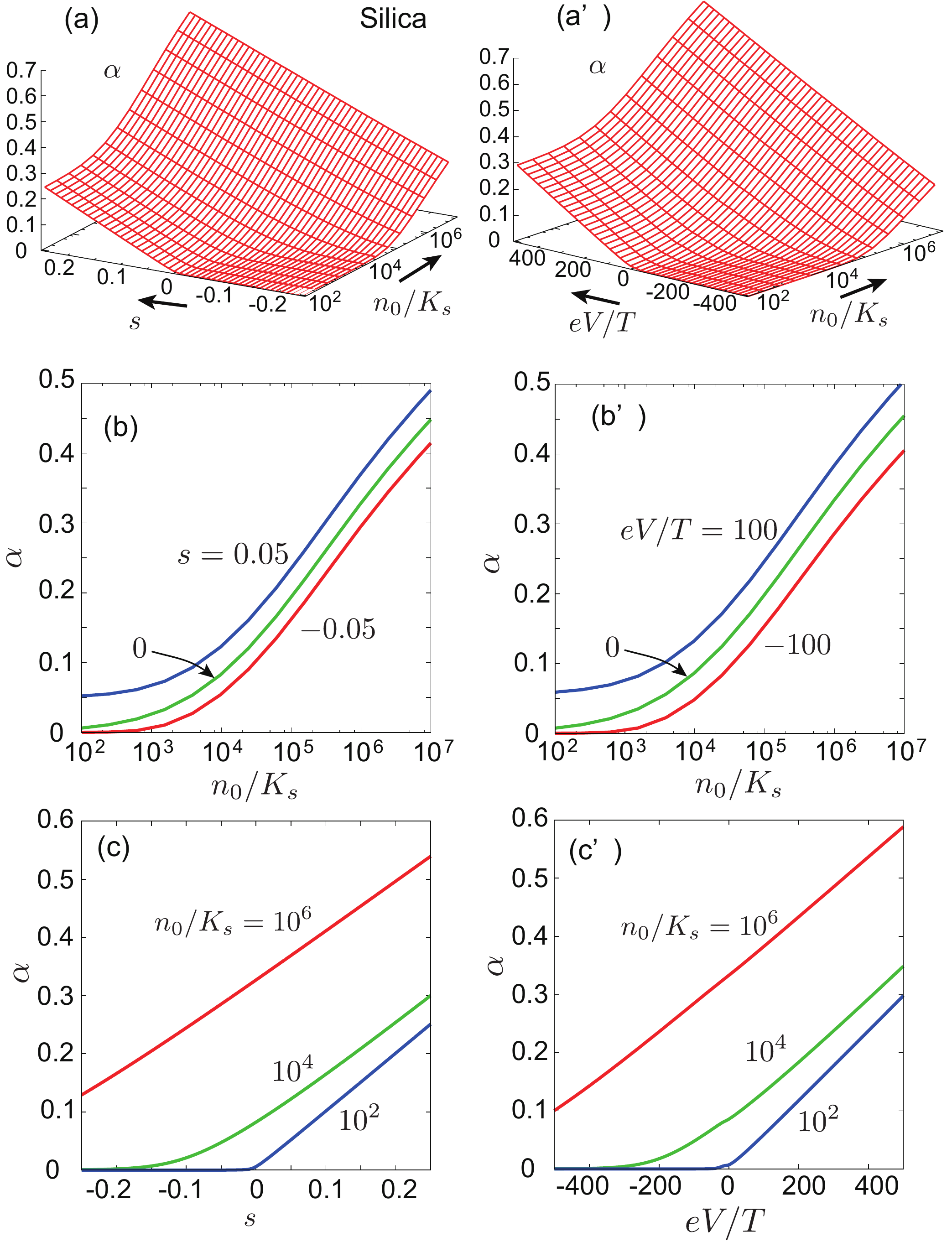}
\caption{Equilibrium $\alpha$ for silica in  charge-controlled case 
(left) and in  potential-controlled case (right). 
Top:  $\alpha$ in the  $s$-$n_0/K_s$ plane 
in (a)  and in the $eV/T$-$n_0/K_s$ plane in (a'), where 
 $s=\sigma_m/e\Gamma_0$.  
Middle: $\alpha$ vs $n_0/K_s$ at 
$s=0$ and $\pm 0.05$ 
   in (b) and  $eV/T=0$ and $\pm 100$ in  (b'), 
where  $n_0/K_s=10^{{\rm pH}-6.7}$ in terms of  the bulk pH. 
Bottom: $\alpha$ vs $s$ in (c) 
and $\alpha$ vs $eV/T$ in (c') for  $n_0/K_s=10^2$, 
$10^4$, and $10^6$, which  are nearly linear.  
 }
\end{figure}

We also   calculate  the integral  $N_{\rm H} 
=\int_o^H dz \nH(z)$ 
using Eqs.(40) and (47). For $\kappa H\gg 1$, 
  we thus obtain 
\bea 
&&\hspace{-1cm} N_{\rm H}/H\nH^0-1
= 2 A_1\alpha/\kappa H
  \nonumber\\
&&\hspace{-1cm}
+{2}[ (A_1^2u^2+1)^{1/2}+ (A_1^2s^2+1)^{1/2}-2]/\kappa H.  
\ena 
The right hand side is the correction 
($\propto H^{-1}$). It   
 can also be equated with   $N_{\rm M}/H\nM^0-1$ for M$^+$  
and $(N_{\rm OH}+ \Gamma_0\alpha)/H\nO^0-1$ for OH$^-$, 
where $N_{\rm M}/H={\bar n}-\ns$ from Eq.(14) 
and $\nM^0=n_0-\ns$ from Eq.(7).  
It follows    $\nH^0\le  N_{\rm H}/H$ for any $u$ and $s$.  
 In particular, if  $u\gg A_1^{-1}$ and $s\gg A_1^{-1}$, 
the right hand side   
is $4A_1 \alpha /\kappa H$ 
 and is negligible only for $H \gg   A_1\alpha\kappa^{-1}$. 
Thus, for   $ \kappa H<  A_1\alpha$, the ion densities 
 $N_i/H$ and  $n_i^0$ 
are largely  different.

\begin{figure}
\includegraphics[scale=0.42]{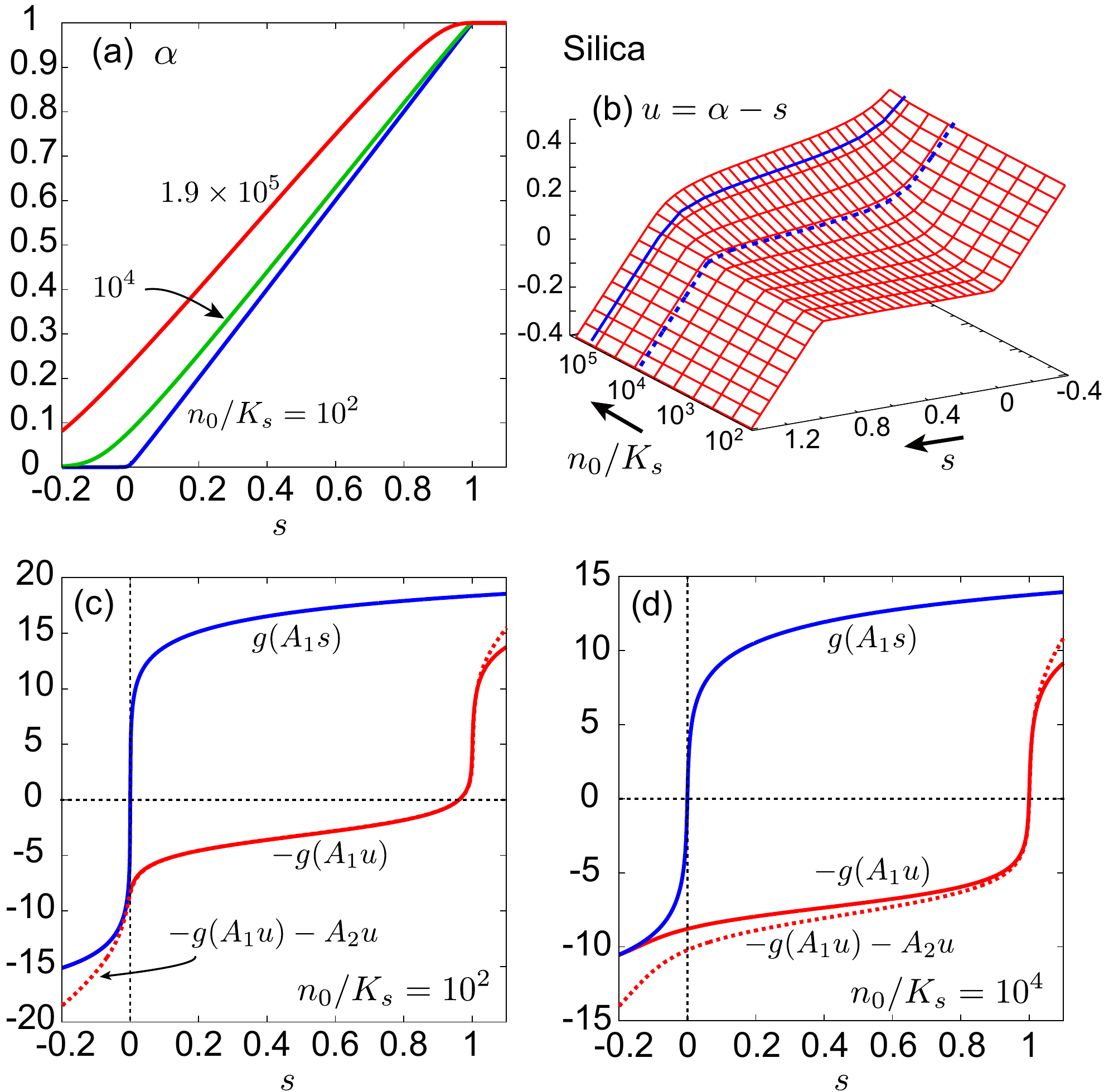}
\caption{ (a)  $\alpha$  vs  $s=\sigma_m/e\Gamma_0$ for 
$n_0/K_s=10^{2}$, $10^4$, and $1.9\times 10^5$ 
in range $[-0.2,1.1]$ for silica, 
 where $\alpha\cong s$. 
(b)  $u= \alpha-s$ in the  $s$-$n_0/K_s$ plane, 
which is very small  for $0<s<1$ and $n_0\ll n_c$. 
The lines (blue) on the surface indicate $n_0/n_c=0.1$ and 1. 
(c) Terms in $eV/T$ in Eq.(54) 
for $n_0/K_s=10^{2}$: 
 $g(A_1 s)$ from the upper diffuse layer, $-g(A_1u)$ from 
the lower diffuse layer, and    $-g(A_1u)-A_2u$ 
including the lower Stern layer contribution.   
(d) Those for $n_0/K_s=10^{4}$.  In (c) and (d), the PB 
equation can be linearized only 
for  $|g(A_1u)|< 1$ at the bottom and  $|g(A_1s)| < 1$ at the top.}
\end{figure}

\subsection{Numerical results on  $\alpha$ }

Use of  Eqs.(40) and (43)  gives 
the ratio of the surface proton density  $\nH(0)$ to the bulk one  $\nH^0$ as  
\bea
&&\hspace{-2cm} 
{\nH(0)}/{ \nH^0} = \exp[g(A_1 u) + A_2u] 
\nonumber\\
&&\hspace{-0.5cm} 
= [({1+A_1^2u^2} )^{1/2}  +A_1 u]^2  e^{A_2 u}, 
\ena 
which depends on $u$ and $n_0$. 
The right hand side increases with increasing $u$ 
being equal to  1 for $u=0$. 
In Eq,(52),  $-u$ denotes the normalized, effective surface 
charge density at $z=0$. Thus, for $u>0$ ($u<0$),  
the protons are more enriched (decreased) 
near the surface than in the bulk.
 From $A_1\gg 1$, 
 ${\nH(0)}/{ \nH^0}$  can be  of order  only in the narrow range 
$|u|< A_1^{-1}$. Outside this range, 
it grows or decays rapidly depending on the sign of $u$ as 
\bea
&&\hspace{-1.2cm}
{\nH(0)}/{ \nH^0} \cong (2A_1u)^2 e^{A_2  u}\gg 1 ~~~~\quad 
(u \gg A_1^{-1}) \nonumber\\ 
&&\cong  (2A_1|u|)^{-2}e^{-A_2 |u|} \ll 1 \quad 
(-u \gg  A_1^{-1}).
\ena  
These changes  occur in the region $z<\ell_{\rm GC}$ 
(see Eq.(57)).

 In the surface chemical equilibrium 
(11),  relevant is  the ratio 
${\nH(0)}/K_s$, which 
 is close to $  (K_{\rm w}/K_s\nO^0)
({\nH(0)}/\nH^0)$ and is calculated from Eq.(59).  
 In the left panel of  Fig.4, it is plotted  
in the $u$-$n_0/K_s$ plane. From Eq.(60), 
 it grows abruptly  from small values $(\ll 1$) 
to large values ($\gg 1$) with increasing $u$ above  0.  
In the right panel, we  show curves of  ${\nH(0)}/K_s $ vs $u$ for 
 several  $n_0/K_s$,
for which  $A_1$ is indeed large ($ 2.5 \times 10^3$ 
for curve (a) 
and  $3.9 \times 10^2$ for curve (e)). 
 For  each point on the surface or the curves in Fig.4 
we can find  the corresponding   equilibrium by  setting   
 $\alpha=[1+ {\nH(0)}/K_s ]^{-1}$ and 
 $\sigma_m= e\Gamma_0 ( \alpha-u)$.  

Substitution of  Eq.(59)  into  Eq.(36) gives $d \Omega (\alpha)/d\alpha$ 
as a function of $\alpha$ for  given  $\sigma_m$ and $n_0$. 
In Fig.5, we plot it and its integral 
$\Delta\Omega =\int_0^\alpha d\alpha (d\Omega/d\alpha)=  
\Omega(\alpha) -\Omega(0)$ as functions of $\alpha$ for four 
$s=\sigma_m/e\Gamma_0$. See Appendix D for the explicit expression 
of $\Delta\Omega$. 
Minimization  of $\Delta\Omega(\alpha)$  yields  
equilibrium $\alpha$ satisfying Eq.(11). 
For $n_0/K_s= 
2.5\times 10^2$ (left),  $d\Omega(\alpha)/d\alpha$ 
grows  abruptly and $\Delta\Omega(\alpha)$ 
has  a cusp-like minimum around  $\alpha\cong s=\sigma_m/e\Gamma_0$ 
with  $0<u\ll 1$ (see Fig.7b). 
For a larger $n_0/K_s=6.3\times 10^4$ (right), 
 $d\Omega(\alpha)/d\alpha$ grows  gradually and 
$\Delta\Omega(\alpha)$ has a broad minimum.

In Fig.6,   we plot the  equilibrium 
$\alpha$   as a function of   
$\sigma_m$ and $n_0$ 
 (left) and as a function of  $V$ and  $n_0$ (right), 
where $\sigma_m$ and $V$ are related by    Eq.(54). 
Salient  results  are as follows. 
(i) We can see close resemblance between 
   the  left  and right panels, which  suggests 
  an approximate  linear relation $V\propto \sigma_m$ (see Sec.IIID). 
In these panels,  $\alpha$ increases with 
increasing $n_0$ in agreement  with the experiment 
\cite{Gisler,Beh4,Yama}.
(ii) In  the middle,   we write   $\alpha$ vs $n_0/K_s$, where  
we fix  $s=\sigma_m/e\Gamma_0$ at $ 0$ and $\pm 0.05$ 
   in (b) and  $eV/T$  at 0 and $\pm 100$ in  (b').  
These  curves exhibit different behaviors with a crossover at 
$n_0/K_s\sim 10^4$. The curve of $\sigma_m=0$ 
exhibits the power-law 
behavior $\alpha \propto n_0^{2/3}$ 
for small $n_0$ (see Eq.(63)).  
(iii) In  the bottom,   at  fixed  $n_0$,  
 we write   $\alpha$ vs  $s=\sigma_m/e\Gamma_0$  in (c) 
and  $\alpha$ vs  $eV/T$ in  (c'), where    $\alpha$ increases 
linearly with increasing  $s$ or $eV/T$. See below for its explanation.

 Note that  the 
  $\alpha$-$n_0$  relation 
  in the 1D geometry  at $s=0$  can  
be used for colloidal particles with large 
radius ($\gg \kappa^{-1}$)\cite{Beh,Beh1,Beh0}. 
At $\sigma_m=0$,   Behrens and Grier \cite{Beh} 
calculated the  curves of $\alpha$ vs pH in agreement with 
Fig.6b.  At the special point 
 $\sigma_m=n_0=0$, 
we find $\alpha= 4.7\times 10^{-4}$ and  
$\sigma_{\rm A}= 
-0.06$ $\mu$C$/$cm$^2$, where the latter  is close to  its experimental value 
 $-0.08$ $\mu$C$/$cm$^2$ for silica colloidal particles 
in pure water\cite{Yama}.

\subsection{Equation for $\alpha$ in nonlinear regime}

In  the nonlinear PB regime  with $ u=\alpha-s 
\gg A_1^{-1}$,    Eqs.(11) and (60) give the following equation of 
 $\alpha$,    
  \bea 
{\alpha u^2}   
e^{A_2  u}/(1-\alpha)&=& K_s \kappa^2
/[(2\pi\ell_B\Gamma_0)^2  \nH^0]\nonumber\\
&=&   ({\nO^0}/n_c)^2 .
\ena
In the second line, we set $\kappa^2/\nH^0= 8\pi \ell_B 
(\nO^0)^2/K_{\rm w}$ and we introduce 
 a crossover ion density   $n_c$  by 
\be 
n_c= (2\pi \ell_B K_{\rm w} /K_s)^{1/2} \Gamma_0 .
\en 
For silica with NaOH, we find 
$n_c= 0.97\times 10^{-2}{\rm mol}/{\rm L}
=1.9 \times 10^5 K_s$ from Eq.(15). 
For surfaces with smaller pK (much larger $K_s$), 
we may add  HCl to increase $\nH^0$ 
(instead of NaOH)  
at densities much larger than $ K_{\rm w}^{1/2}$; then,  
 the right hand side of Eq.(61)  becomes $K_{\rm w}/n_c^2= K_s / 
[2\pi \ell_B \Gamma_0^2]$.

We note that  Eq.(61) reproduces the curves of $|s|\ll 1$  in Fig.6b, 
where we can see  $\alpha\ll 1$  for $n_0 \ll n_c$. 
 In particular,  for $s=0$ and $ n_0\ll n_c$, Eq.(61) gives 
\be
\alpha\cong 
\alpha_c = ( n_{\rm OH}^0 /n_c)^{2/3}
= \alpha_{c0} ( n_{\rm OH}^0 /K_{\rm w}^{1/2})^{2/3}. 
\en 
In pure water, $\alpha_c$ tends to the following, 
\be 
\alpha_{c0} = ( K_{\rm w}^{1/2}/n_c)^{2/3}=( K_s/2\pi \ell_B)^{1/3}
/\Gamma_0^{2/3}, 
\en 
For silica, we find $\alpha_{c0}= 
4.7\times 10^{-4}= 5.4/A_1$ and  $U(d_0) =-7.0$, so 
 we are  already in the nonlinear PB 
regime    at  $\sigma_m=n_0=0$. 
In terms of the bulk pH, we obtain  $\log_{10}(\alpha/\alpha_{c0})=
{2({\rm pH} -7)/3}$.  Note that the relation (59) applies to 
large   colloidal particles  in the nonlinear PB  regime. 
On the other hand, for  $n_0>n_c$, $u$ increases 
and  the factor $e^{A_2  u}$  
  becomes  important.

Let us assume $ n_0\ll n_c$ for $n_c \gg K_{\rm w}^{1/2}$.  
In Fig.7a,   we confirm  the  behavior 
$\alpha\cong s$ in the range $ 0<s<1$, 
while  we have $\alpha\cong 0$ for $s<0$ and $\alpha\cong 1$ for $s>1$. 
In (b), this  behavior can be seen on a flat  part of the surface of 
$u=\alpha-s$  in the  $s$-$n_0/K_s$ plane. 
Here, {\it self-regulated ionization} is  realized, 
 where   the surface charge 
density $ \sigma_{\rm A}$ 
due to deplotonation   
nearly   cancels the applied surface charge 
density $\sigma_m$.  To be more precise,     Eq.(61)  yields   
\be 
u=\alpha-s  \cong (1-s)^{1/2}s^{-1/2} n_{\rm OH}^0/n_c
\en 
which is  much smaller than $s(\cong \alpha)$ for 
 $s\gg \alpha_c $ with $\alpha_c$ being  defined by  Eq.(63). If 
 $|s|< \alpha_c$, $\alpha$ becomes of order $ \alpha_c$,  
which is consistent with  Eq.(63).  On the other hand, if 
$n_c/K_{\rm w}^{1/2}$  is not large 
  (with smaller  $\Gamma_0/K_s^{1/2}$), $u$ remains not small 
for any $n_0$. 

Finally, we need to require 
$A_1 u\gg 1$  self-consistently, 
which has been assumed in  setting  up Eq.(61). For $|s|\gg  \alpha_c$, 
use of  Eqs.(50) and (64) gives   
\be 
A_1 u \cong (1-s)^{1/2}s^{-1/2}  
(n_{\rm OH}^0K_s/4K_{\rm w})^{1/2}.
\en  
If $n_{\rm OH}^0 \gs   K_{\rm w}/K_s$, 
$A_1 u$  surely exceeds 1  for $s$ not very close to 1, 
including the point  $s=n_0=0$.
Thus, it generally follows  the self-regulation of surface ionization 
 for  $n_0\ll  n_c$ and $ 0<\sigma_m<e\Gamma_0.$

\begin{figure}[tbp]
\includegraphics[scale=0.4]{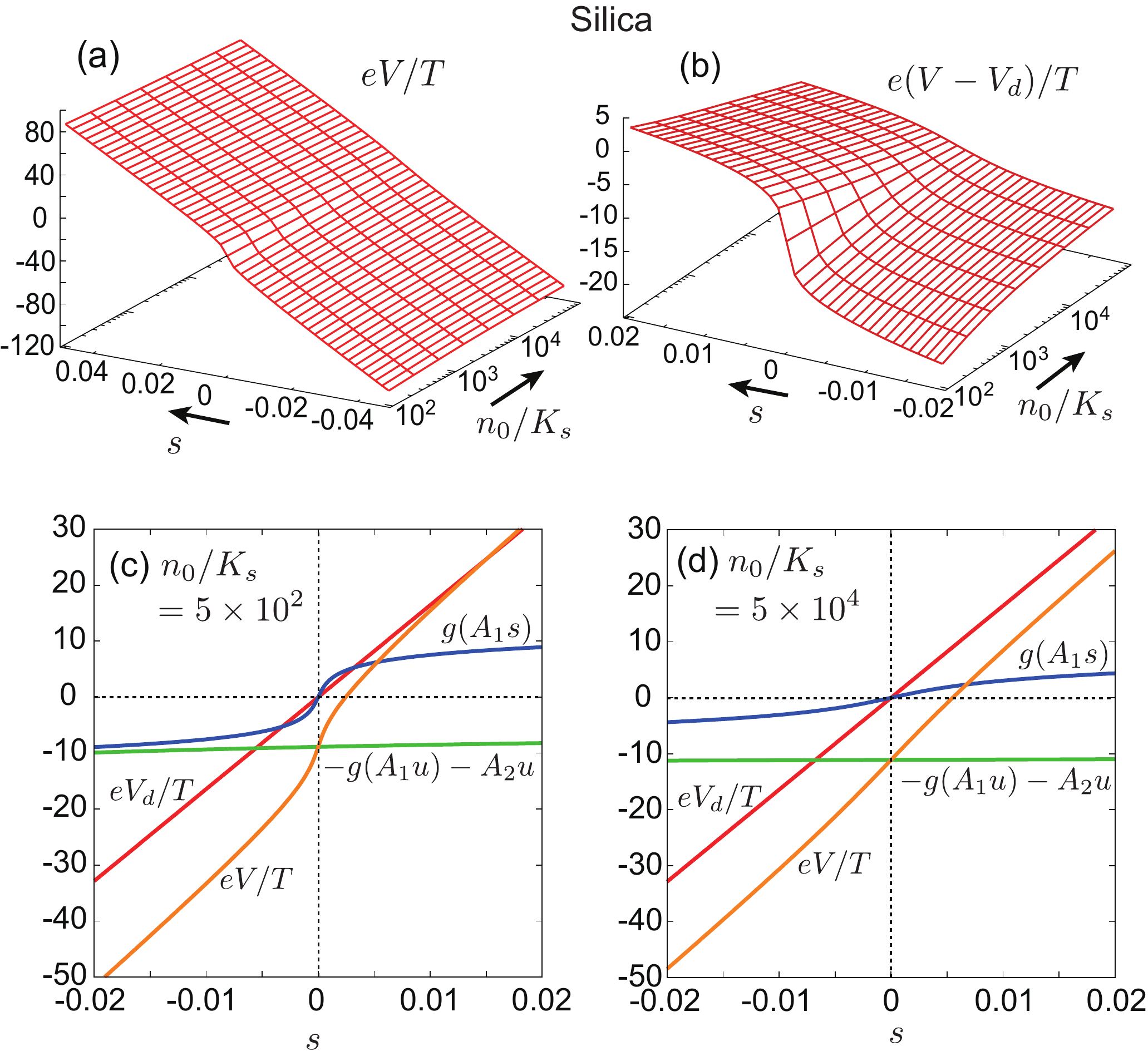}
\caption{ Potential contributions in Eq.(54) (silica). 
Top: (a)    $eV/T$  ($|s|<0.05)$ 
and  (b)  $e(V-V_d)/T$  ($|s|<0.02)$ 
 in the  $s$-$n_0/K_s$ plane, where  
  $s=\sigma_m/e\Gamma_0$ and $V_d=\sigma_m/C'$. 
Bottom: Plotted are  $-g(A_1 u)-A_2u$, 
$g(A_1s)$, $V_d$, and  their sum  $V$ as functions of  $s$, 
where  $n_0/K_s$ is (c) $5\times 10^2$ and (d) $5\times 10^4$. 
 See (c) and (d) of Fig.7 
for expanded  plots of  $-g(A_1 u)$,  $-g(A_1 u)-A_2u$, 
and $g(A_1s)$.}
\end{figure}

\subsection{Potential difference $V$ between electrodes}

In Appendix A, we will explain experimental setups 
at fixed $V$ and  $\sigma_m$, so we should 
 compare   the results  from  these two 
boundary conditions.   So far we have found that 
the  right and left panels in Fig.6 look similar. 

In Eq.(54) the contributions to  $eV/T$ 
are  written explicitly. In the bottom panels of  Fig.7, 
we examine them 
 in the range $-0.2<s=\sigma_m/e\Gamma_0<1.1$ as in the upper panels. 
Here,    $g(A_1 s)$ is 
from the upper diffuse layer, $-g(A_1u)$ is from 
the lower diffuse layer, and    $-g(A_1u)-A_2u$ 
includes  the lower Stern layer contribution, where 
  $n_0/K_s$ is $10^{2}$ in (c) 
and   $10^{4}$ in (d). For  positive $s$ not close to 0, 
  the contribution from 
the lower diffusive layer is  smaller 
than that from the upper one in magnitude,  
which is consistent with  Fig.7b. 
Here, for  $u\gg A_1^{-1}$ and $s\gg A_1^{-1}$, 
we have  $U(d_0)\cong 
-2 \ln (2A_1 u)$ and $U(H-d_H) \cong -2\ln (2A_1s)$.  
We  are  thus in the nonlinear PB regime  for most $s$. 
The contribution $-A_2u$ from   the lower Stern layer 
is appreciable  for   $s<0$  and $0<s\ll 1$ 
or for $\alpha\ll 1$.

In Fig.8, we display (a) $eV/T$ and 
(b) $e(V-V_d)/T$  in the $s$-$n_0/K_s$ plane 
with $V_d=\sigma_m/C'$, 
where  the  surfaces are rather   flat 
for not very small $s$. We  also compare  $eV/T$,   $eV_d/T$, 
$g(A_1 s)$, and    $-g(A_1u)-A_2u$  
in the narrow range $|s|<0.02$ 
for $n_0/K_s=5\times 10^2$ in (c) and $10^4$ in (d).  
We recognize that 
 $V$ is close to $V_d$ except for small $\sigma_m $ even for  
our choice $\ell=1.05$ nm.  With further increasing $\ell$, 
the film contribution $\sigma_m/C_d$ becomes more dominant in $V$,  
 as in electrowetting experiments \cite{Mug,Andel1}.

\section{Ionization on Carboxyl-bearing  surface at large wall separation }
\begin{figure}[tbp]
\includegraphics[scale=0.4]{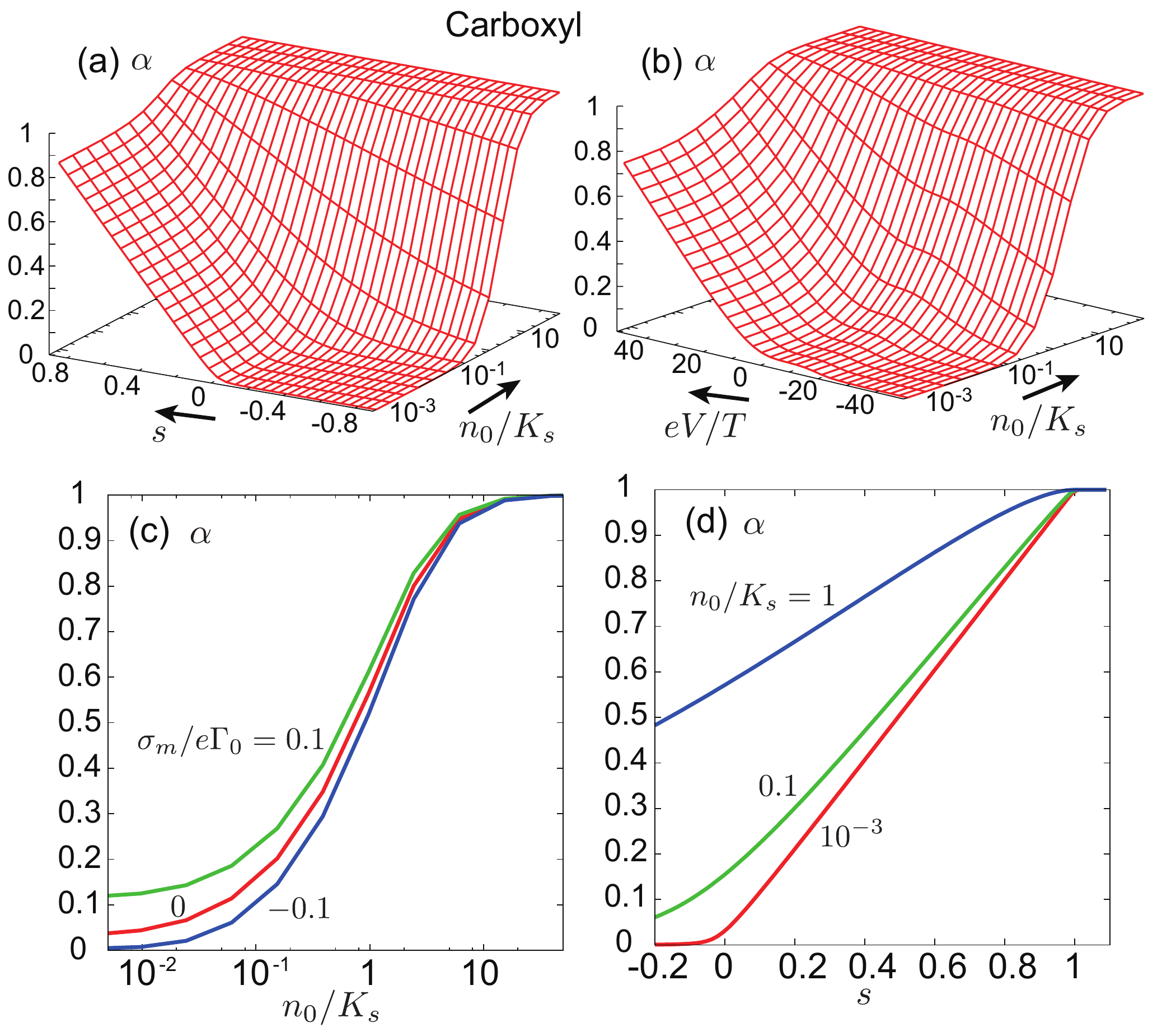}
\caption{Ionization on  carboxyl-bearing  surface. 
(a) $\alpha$ in the  $s$-$n_0/K_s$ plane 
and (b) $\alpha$ in the  $eV/T$-$n_0/K_s$ plane,  
 where $s=\sigma_m/e\Gamma_0$.     
(c) $\alpha$ vs $n_0/K_ \sigma$ at 
$s=0$ and $\pm 0.1$. (d) 
$\alpha$ vs  $s$ for 
$n_0/K_s=10^{-3}$, $0.1$, and $1$ in range $[-0.2,1.1]$.
 }
\end{figure}

The carboxyl surface groups    undergo  
the dissociation (COOH  $\rightleftarrows$
COO$^-$ + H$^+$) at  pK$= 4.9$ with a 
much smaller   $\Gamma_0$ in water.  In their analysis, 
Behrens {\it et al}\cite{Beh,Beh0,Beh1} used two values, 
 $\Gamma_0= 0.574$ and 0.250 nm$^{-2}$,   
for carboxyl-bearing surfaces.   
These pK and $\Gamma_0$  are  very different from 
those in Eq.(15) for  silica.  
It is worth noting that Aoki {\it et al.}\cite{Aoki}  fabricated a 
 carboxyl functionalized latex film 
with coalescence of latex particles on a Pt electrode.
 Analysis has also been made on  other 
surfaces such as 
iron oxide ones  \cite{BehR,Beh0,Beh}, 
which can be  positively charged at low pH 
due to protonation.

In this section, we examine 
 the equilibrium ionization on a carboxyl-bearing film 
in the 1D geometry in Fig.1 
with addition of NaOH in applied  field. We set   
\bea 
&&\hspace{-1cm}
K_s= 10^{-4.9}{\rm mol}/{\rm L} 
= 7.5 \times 10^{-6}/{\rm  nm}^{3}   ,\nonumber\\
&&\hspace{-1cm}
\Gamma_0=  0,25/{\rm nm}^2 \quad ({\rm carboxyl}{\rm -bearing}~{\rm surface}).
\ena 
Here,  $K_s$ is much larger than $K_{\rm w}^{1/2}$ by 
$10^{2.1}=126$. According to Behrens {\it et al.}\cite{Beh,Beh0,Beh1}, 
the Stern capacitance  at carboxyl-bearing  surfaces 
is larger than that for silica oxide  surfaces 
such that  its presence itself can be neglected. 
 In fact,  for $C_0=10$ F$/$m$^2$,  $A_2$ in Eq.(51) is of order 0.1 
  so  $e^{A_2u}\cong 1 $ in Eq.(59) for $|u|\ls 1$.  
The other parameters are the same as in the silica case in Sec.III.

\begin{figure}[tbp]
\includegraphics[scale=0.47]{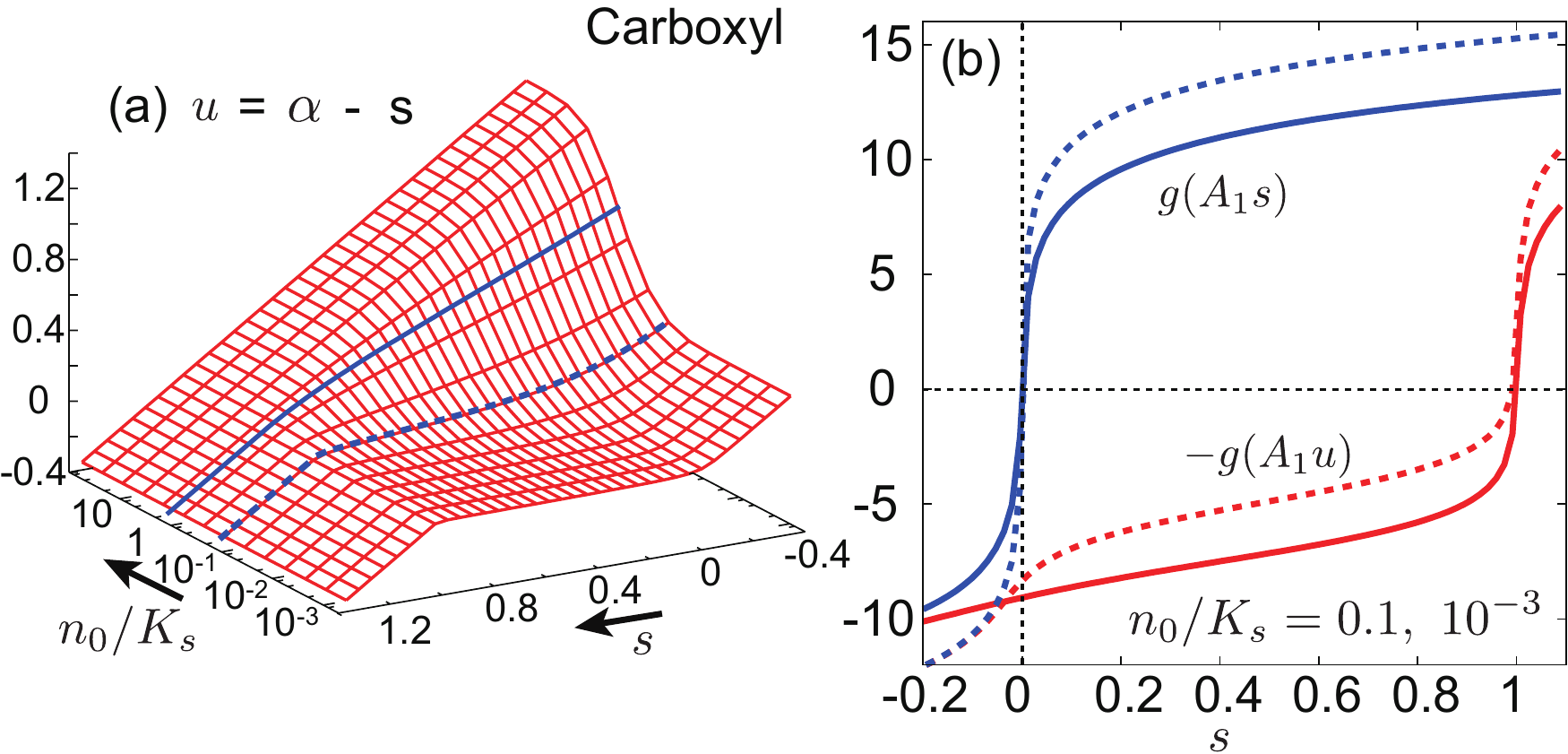}
\caption{ 
(a) $u= \alpha-s$ in the  $s$-$n_0/K_s$ plane 
for carboxyl-bearing  surface, 
which is very small  for $0<s<1$ and $n_0<0.1n_c
\sim K_s$. The lines (blue) on the surface 
indicate $n_0/n_c=0.1$ and 1.  
(b) Terms in $eV/T$ in Eq.(54): 
 $g(A_1 s)$ from the upper diffuse layer and $-g(A_1u)$ from 
the lower diffuse layer. 
 for $n_0/K_s=0.1$ (bold line)  and $10^{-3}$ (dotted line).
}
\end{figure}

We use the PB  theory in the previous section 
 for large $H$. The parameter $A_1$ in Eq.(50) 
becomes   $A_1=1.5\times 10^3  (K_s/\nO^0)^{1/2}$.  
In  Eq.(61) 
we neglect the Stern factor $e^{A_2 u}$ and assume 
 $ u=\alpha-s\gg A_1^{-1}$ to obtain    
\be
{\alpha u^2}   /(1-\alpha) \cong  (\nO^0/n_c)^2, 
\en 
where $ n_c = 1.9 \times 10^{-5}  {\rm mol}/{\rm L}= 
1.5K_s$ from Eq.(67). It is  smaller than that for silica by $500^{-1}$.   
At the special point $n_0=\sigma_m=0$,  
our numerical analysis indicates 
 $U(0)=-8.32$ and  $\alpha_{c0}= 
0.0a30= 32/A_1$, where the latter nearly coincides with  
$\alpha_{c0}$ in Eq.(64). 
We are in the nonlinear PB 
regime at this point.

In Fig.9,  we show $\alpha$ in (a) the  $s$-$n_0/K_s$ plane 
 and in (b) the  $eV/T$-$n_0/K_s$ plane,   as in (a) and (a') 
of Fig.6.  Here,   $\log_{10}(n_0/K_)= {\rm pH}-9.1$.  
These look similar as in Fig.6, indicating the linear relation 
$V\cong \sigma/C'$ except for small $\sigma_m$. 
 In (c),  $\alpha$ vs $n_0/K_s$ is plotted for $s=0$ and 
$\pm 0.1$, as in Fig.6b. 
Here, $\alpha$ approaches 1 for $n_0 \sim 6K_s 
\sim 4n_c$.   The curve at $\sigma_m=0$ coincides with that 
calculated by Behrens and Grier \cite{Beh}. 
In (d), we display 
$\alpha$ vs  $s$ for 
$n_0/K_s=10^{-3}$, $0.1$, and $1$ in 
the range $-0.2,s<1.1$, as in Fig.7a. 
We can  again find the linear behavior $\alpha\cong s$ 
for $0<s<1$ for $n_0\ll n_c$.

In Fig.10, displayed is (a) $u= \alpha-s$ in the  $s$-$n_0/K_s$ plane 
for carboxyl-bearing  surface, 
which is very small  for $0<s<1$ and $n_0<0.1n_c\sim 
/K_s$, as in Fig.7b. The lines (blue) on the surface 
indicate $n_0/n_c=0.1$ and 1.  
(b) Terms in $eV/T$ in Eq.(54) 
for $n_0/K_s=10^{2}$: 
 $g(A_1 s)$ from the upper diffuse layer, $-g(A_1u)$ from 
the lower diffuse layer. 
 for $n_0/K_s=0.1$ (bold line)  and $10^{-3}$ (dotted line).

\section{Ionization and disjoining pressure 
at small wall separation}

The disjoining pressure $\Pi_d$ in Eq.(46) 
is a measurable qunatity\cite{Is,Is1,Ga,Zhao}. It 
has been calculated between  ionizable surfaces  at 
 small  separation  \cite{Par,Chan,AndelEPL}. 
Here, attaching a reservoir, 
we examine  how  $\Pi_d$ and $\alpha$ depend on $H$ and $\sigma_m$ 
with  NaOH added at a density $n_0$. 
We consider  the  Stern layers, so 
we define the effective cell thickness  by
\be 
H'= H-(d_0+d_H), 
\en 
where $d_0$ and $d_H$ are 
of order $5 {\rm \AA}$. There is a sizable range of 
$H'<\kappa^{-1}$ for not large  $n_0$. 
Using  Eqs.(11) and (40),   
we  integrated the PB equation in the region  $0<z-d_0<H'$ 
for  each given $n_0$ and $\sigma_m$.  The parameters   
are those for  silica oxide  surfaces in Figs.11-16, 
but those for  carboxyl-bearing surfaces 
are also used in Fig.17.  
For simplicity, we neglect the van der Waals interaction\cite{Hunter,Is,Butt}.

\subsection{Results without applied  field ($s=0$)}

First, we assume no applied field $(s=0)$. 
  Because $dU/dz=0$ at $z=H-d_H$, 
our $H'$ corresponds to a half of the cell
 thickness in the symmetric case\cite{Par,AndelEPL}. 
For thick cells with  $H'>   \kappa^{-1}$, it follows  
a well-known result \cite{Andelbook,Is,Ohshima}, 
\be 
\Pi_d \cong 64T\nO^0 \exp(-2\kappa H').
\en

\begin{figure}[tbp]
\includegraphics[scale=0.43]{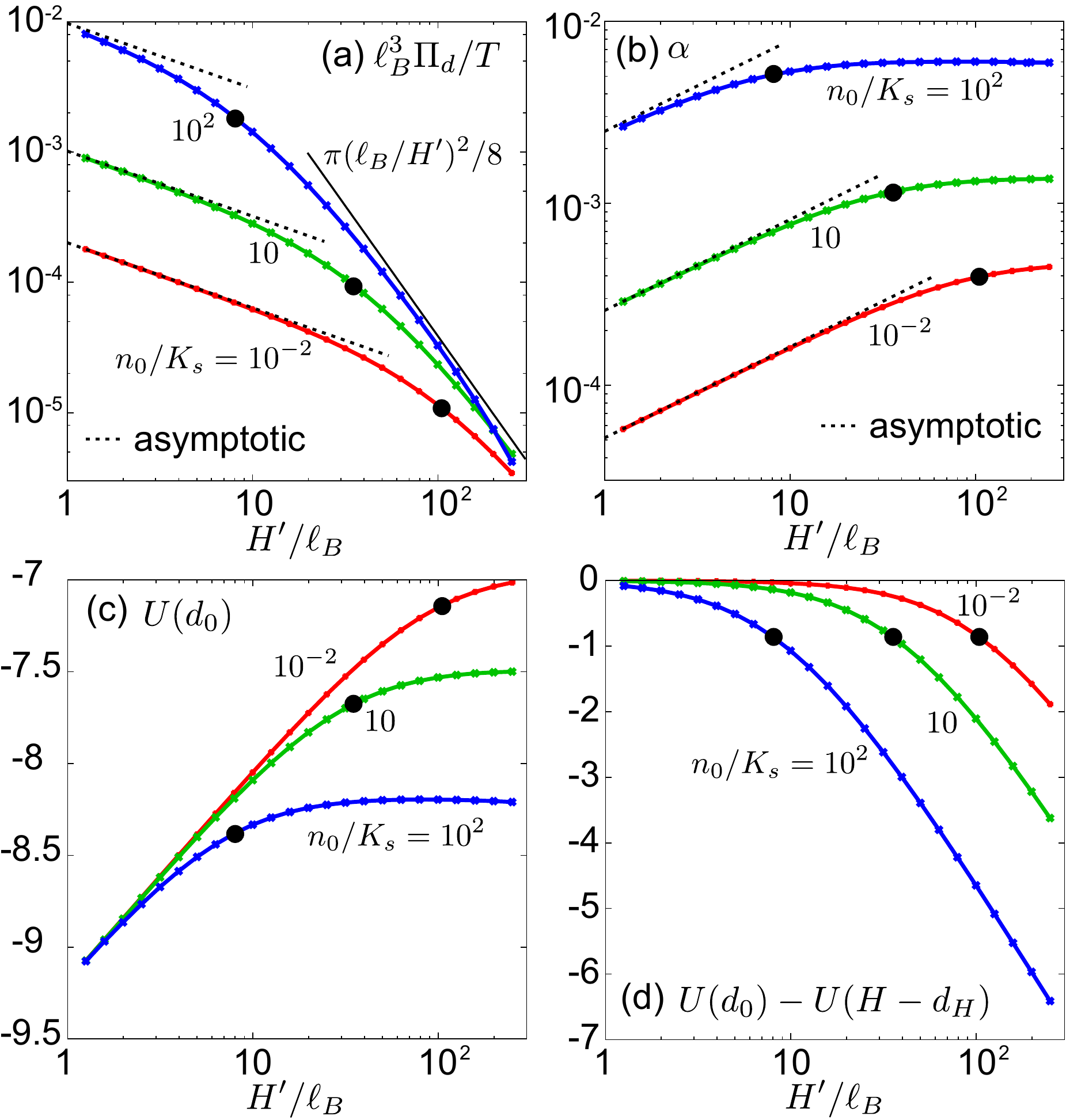}
\caption{ Results for silica in the range 
$1<H'/\ell_B<300$  ($\ell_B=7~{\rm \AA}$)  
 without applied field ($s=0$) 
for $n_0/K_s=10^{-2}, 10$, and $10^2$. Plotted  are 
(a)  $\ell_B^3 \Pi_d/T$, (b) $\alpha$,  
(c) $U(d_0)$, and (d) $U(d_0)-U(H-d_H)$. Points at which 
$H'=\ell_{\rm GC}$ are   marked by  $\bullet$ on each curve. 
In (a) and (b),   formulae  (73), (74), and (78) 
are written as guides of eyes.  In (a)-(d), 
Eqs.(71)-(75)  hold for  $H'\ll \ell_{\rm GC}$ 
and Eqs.(76)-(78)  for $H'\gg \ell_{\rm GC}$.
 }
\end{figure}

Ninham and Parsegian \cite{Par} 
derived  the power-law behavior  $\Pi_d \propto  H^{-1/2}$ 
as  $ H\to 0$ between  symmetric  
ionizable walls. 
 From  analysis in Appendix E, their asymptotic 
 behavior appears  for 
$H'\ll \ell_{\rm GC}$
in the case $ A_1^{-1} \ll  \alpha \ll  1. 
$ 
Here,  $\ell_{\rm GC}=(\kappa A_1 \alpha)^{-1}$  is the Gouy-Chapman length 
in Eq.(56)   shorter 
than  $\kappa^{-1}$ in the nonlinear PB regime. 
  The normalized potential values 
$U_0= U(d_0) $ and $U_H=U(H-d_H)$ at the two ends are given by 
\bea 
U_H-U_0 &\cong&  A_1\alpha \kappa H'= H'/\ell_{\rm GC}  \ll 1 , \\ 
\exp({-U_0})  &\cong&  4A_1 \alpha/\kappa H'\gg 1. 
\ena 
From Eq.(11) and (46),  $\Pi_d$ and $\alpha$  behave as    
\bea
&&\hspace{-1.7cm} \Pi_d \cong     T\nO^0   e^{-U_H} 
  \cong   T\nO^0  (\Gamma_0 K_s/K_{\rm w} H')^{1/2},\\
&&\hspace{-1.6cm}\alpha \cong  (K_s H' /\Gamma_0 K_{\rm w})^{1/2}\nO^0 ,
\ena 
where  we set $\nH^0=K_{\rm w}/\nO^0$. 
 With    other added ions, however, 
 Eqs.(73) and (74) should be changed appropriately 
(see the comments below Eqs.(42) and  (46))\cite{Par,AndelEPL}. 
In addition, the inequality  $\alpha \ll 1$ holds  for 
\be 
H' \ll (\Gamma_0/ K_s) (K_{\rm w}^{1/2}/\nO^0)^2. 
\en 
For silica oxide  surfaces, we have  $\Gamma_0/K_s\sim 10^8$ nm, 
so Eq.(76)  can well be satisfied 
together with Eq.(71).

On the other hand, when $ \ell_{\rm GC}\ll H'\ll \kappa^{-1}$ 
in the nonlinear PB regime, 
we find  another  regime (the Gouy-Chapman regime\cite{Andelbook}), 
where  $U_0$ and $\alpha$ remain nearly at constants  
 but $U_H$ strongly depends on $H'$. That is,  
\bea 
&&\exp({-U_0}) \cong (2A_1\alpha)^2, \\
&&\exp({-U_H}) \cong (\pi/\kappa H' )^2.
\ena 
Thus, $e^{U_H-U_0}\sim (H'/\ell_{\rm GC})^2\gg 1$. 
The $\Pi_d$ behaves as \cite{Andelbook}
\be 
\Pi_d \cong  T \nO^0 e^{-U_H} \cong \pi T/[2 \ell_B(2 H')^2],   
\en  
which is independent of $n_0$ and $\alpha$. 
In the previous theories\cite{Par,AndelEPL},  
the  behavior (78) was not found,   
because they adopted  parameters yielding 
 $\ell_{\rm GC}\gs \kappa^{-1}$ 
with a salt added (see the last pragraph of Sec.IIIA).

\begin{figure}[tbp]
\includegraphics[scale=0.43]{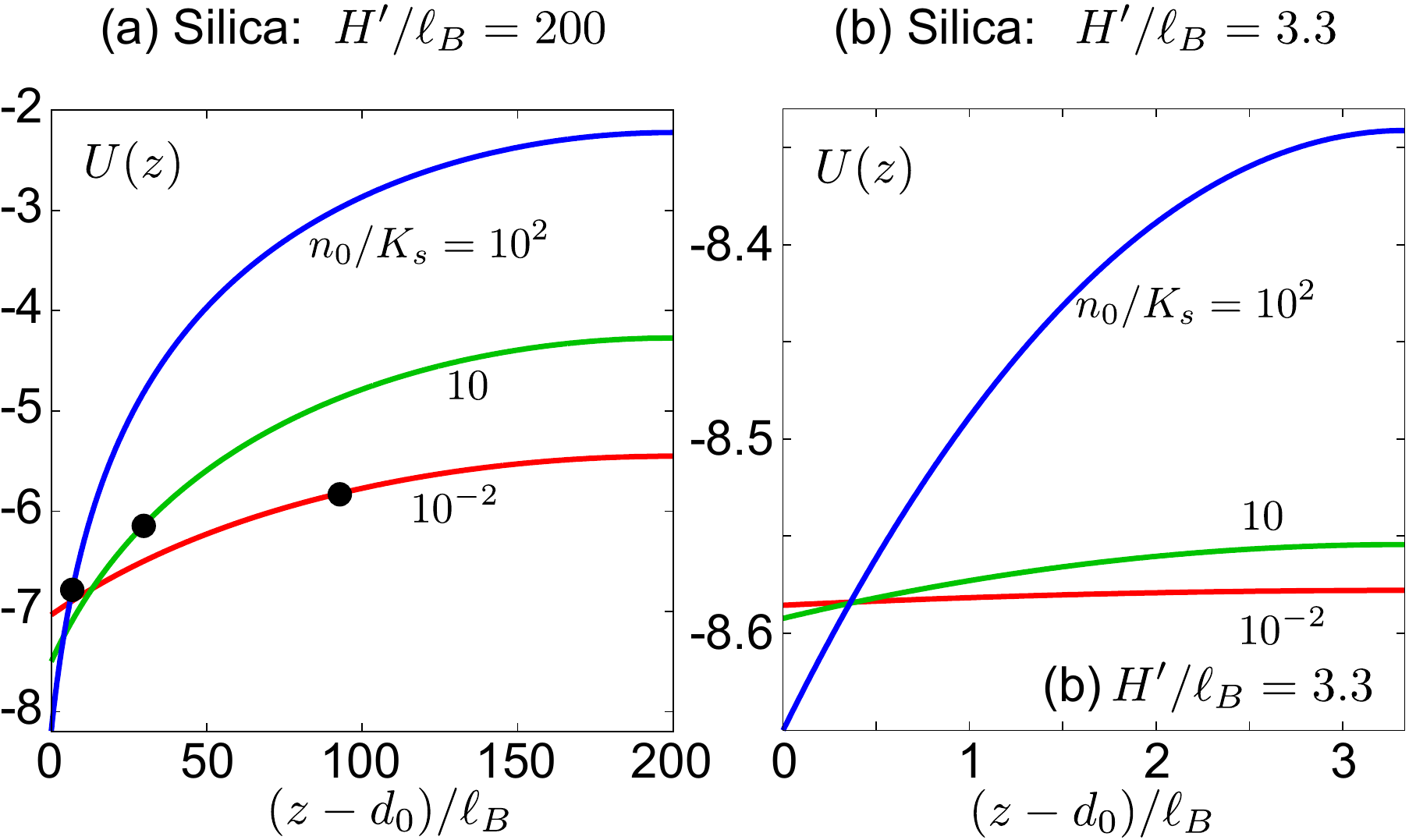}
\caption{Profiles of $U(z)$ in the range $0<z-d_0<H'$ 
 for $s=0$, where a silica oxide surface is at $z=0$. 
 For   $n_0/K_s=10^{-2}, 10$, and $10^2$, 
 $H'$ is  (a)  $200\ell_B$ with $\ell_{\rm GC}< H'<\kappa^{-1}$ 
and  is  (b) 3.3$\ell_B$ with  $H'<\ell_{\rm GC}$. 
}
\end{figure}

In Fig.11, we show  (a) $ \Pi_d$,  (b) $\alpha$,  
(c) $U_0$, and (d) $U_0-U_H$ 
as functions of $H'$ in the range $\ell_B<H'<300\ell_B< \kappa^{-1}$ 
 without applied field  ($s=0$) for 
silica oxide  surfaces. We set 
  $n_0/K_s=10^{-2}, 10$, and $10^2$, where 
 $\kappa\ell_B\times 10^3$ 
is  $0.72$,  $1.6$,  and $5.0$ 
and  $A_1\times 10^{-3}$ is $34$,  $15$,  and $4.9$, 
respectively. In these cases,  $A_1\alpha$ remains 
considerably larger than 1 for $H'>\ell_B=7~{\rm \AA}$  from Eq.(63). 
We can clearly see the crossover 
between  the  two power-law regimes  at $H'= \ell_{\rm GC}$ as follows. 
In (a), $\Pi_d$ behaves as in Eqs.(73) and (78). 
In (b),  $\alpha$ decays as $\sqrt{H'}$ for small $H'$ and 
is  a small constant for larger $H'$. 
In (c), $U_0$ satisfies Eqs.(72) and (76)  in the range $[-9,-7]$,  
tending to be independent of $n_0$ for $H'\ll \ell_{\rm GC}$. 
In (d), $U_0-U_H$ behaves 
in accord with Eqs.(71) and (77).

In Fig.12, we display  $U(z)$ 
without applied field for $n_0/K_s=10^{-2}, 10$, and $10^2$. 
In  (a), we set  $H'=200\ell_B <\kappa^{-1}$; then, 
$H'> \ell_{\rm GC}$ 
with $\ell_{\rm GC}/\ell_B=6.7$,  $30$, and $93$, respectively.  
In (b), we set  $H'=3.3 \ell_B$ to realize  
$H'<\ell_{\rm GC}$ with 
$\ell_{\rm GC}/\ell_B=440,$  $87$, and $ 10$, where 
the diffuse layers at the bottom and top walls 
largely overlap,

\subsection{Results with applied field}

Next, we apply electric field by varying $s=\sigma_m/e\Gamma_0$ 
for $\kappa H'<1$. In Fig.13,  $ \Pi_d$ 
is displayed as a function of  $ H'$ and $s$. 
We fix  $n_0/K_s $  at  10 and  $10^3$, where 
$A_1$ is $1.5\times 10^4$ and $1.5\times 10^3$, respectively. 
The $ \Pi_d$   is  positive around the line of $s=0$ 
 but is mostly negative. 
This is natural because the  surface charge densities  
at the two ends  have  the same sign  only for 
$0<s<\alpha$. If $\kappa H'\ll 1$,  we find \cite{comment2}   
 that  $\Pi_d>0$ holds only for $|s|\ls \alpha$, 
where $\alpha$ is the degree of ionization at $s=0$. 
 This width of $s$ is  of order 
$\alpha\sim  10^{-3}$ in (a) and 
$10^{-2}$ in (b) from Eq.(63). 
In addition, for not small $|s|$,  $\Pi_d$ drops to negative values 
 with large amplitude with decreasing  $H'$ 
due to partial screening (see below).

\begin{figure}[tbp]
\includegraphics[scale=0.4]{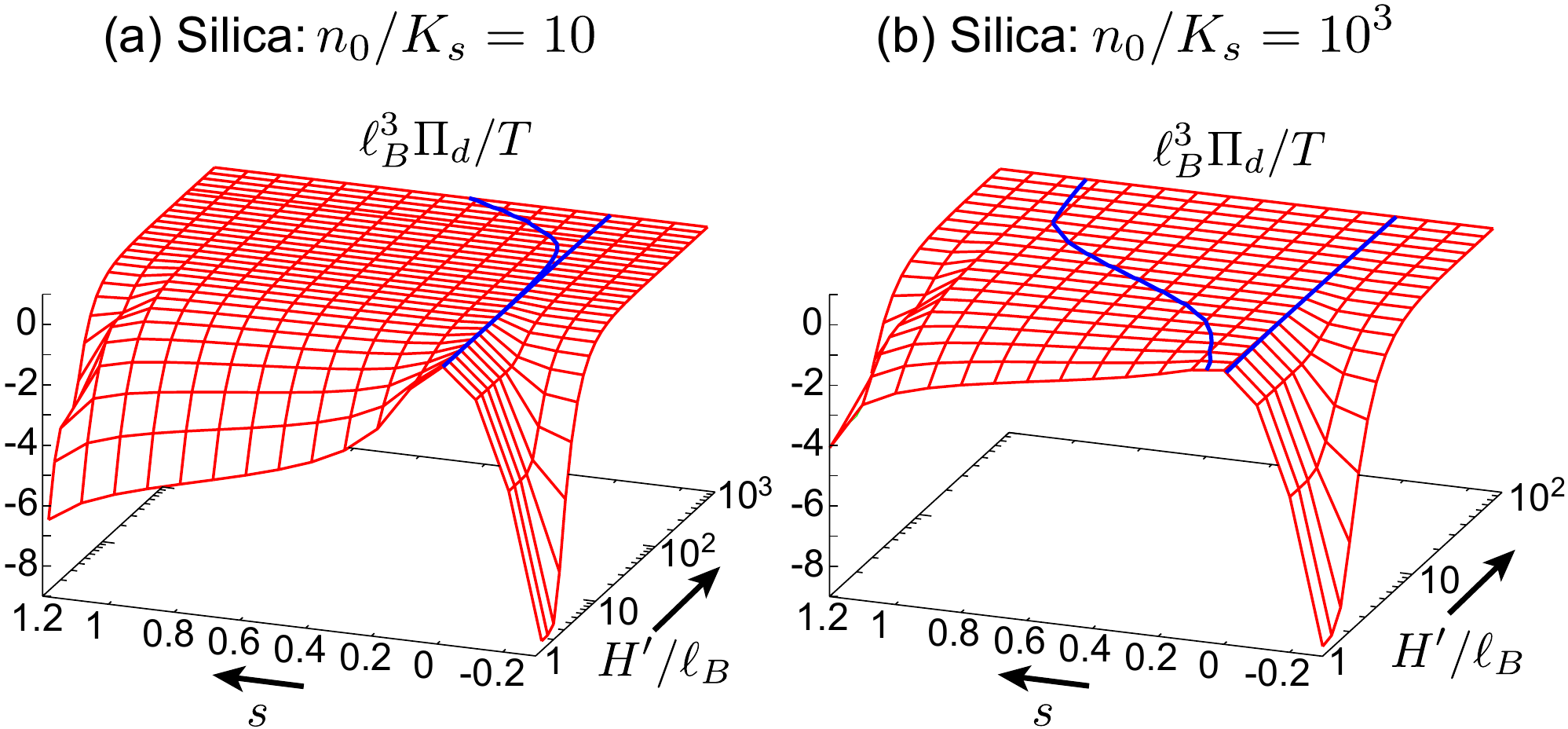}
\caption{ $\ell_B^3 \Pi_d/T$ 
in the $H'/\ell_B$-$s$ plane for small  $ H'$ (silica), 
where $n_0/K_s $ is (a) 10 and (b) $10^3$. 
Lines of $\Pi_d=0$ (blue) are written on 
the surfaces.  Area of $\Pi_d>0$ between the  two lines 
becomes  narrower as $H'$ is decreased 
\cite{comment2} but is widened 
as $n_0$ is incresaed. For relatively large 
 $s$, $\Pi_d$ drops to negative values with large 
amplitude due to partial screening.  
}
\end{figure}

\begin{figure}[tbp]
\includegraphics[scale=0.4]{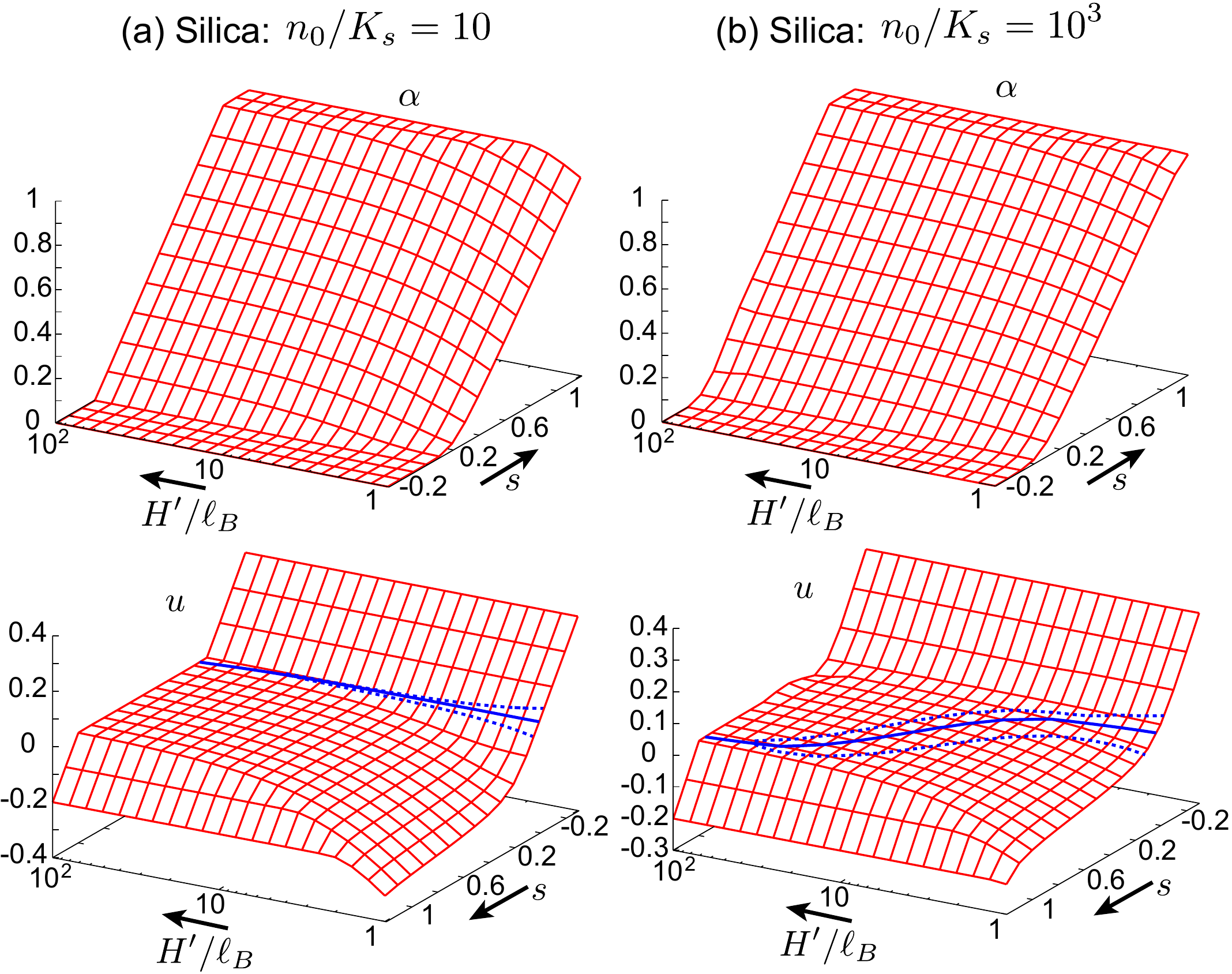}
\caption{ $\alpha$ (top) and  $u= \alpha-s$ (bottom) in 
the $ H'/\ell_B$-$s$ plane  for silica, 
where $n_0/K_s $ is (a) 10 (left) 
and (b) $10^3$ (right). 
For not small $s~(>0)$, 
 $\alpha$ is insensitive to $H'$ 
larger than $5\ell_B$. 
Lines  of $u= (A_1 H')^{-1}>0$, 
$u=0$, and  $u= -(A_1 H')^{-1}<0$ (from above)  are   written on 
the  surfaces  of $u$ (blue lines).  
Below (above) the middle blue line ($u=0)$,  
 $u$ is negative (positive).
}
\end{figure}

In Fig.14, we plot $\alpha$ and  $u= \alpha-s$ 
for   $1< H'/\ell_B<10^2$ 
and $-0.2<s<1.2$, where 
 $n_0/K_s=10  $ (left)  and  $10^3$ (right). 
For   $H'\gs 5\ell_B$, 
  $\alpha$  is nearly independent of $H'$ in the upper panels and  
 $u$ is  very small for $0<s\ls 1$ in the bottom panels.  
Thus,  the self-regulation (found for  $H'\gg \kappa^{-1}$ 
in Sec.III) is  operative  even for  small  $H'(\gs 5\ell_B)$. 
We also write the lines of $u=0$  
and $\pm (A_1 H'\kappa)^{-1}$. We have  $H'> \ell_{\rm GC}$ 
outside  them.  These three  lines are closely located 
because of large $A_1$, which is $1.5\times 10^4$ (left) and 
$1.5\times 10^3$ (right). 
Indeed, on the flat area in the left, 
we have  $-0.01<u<-0.001$ and $A_1|u|> 15$.

\begin{figure}[tbp]
\includegraphics[scale=0.44]{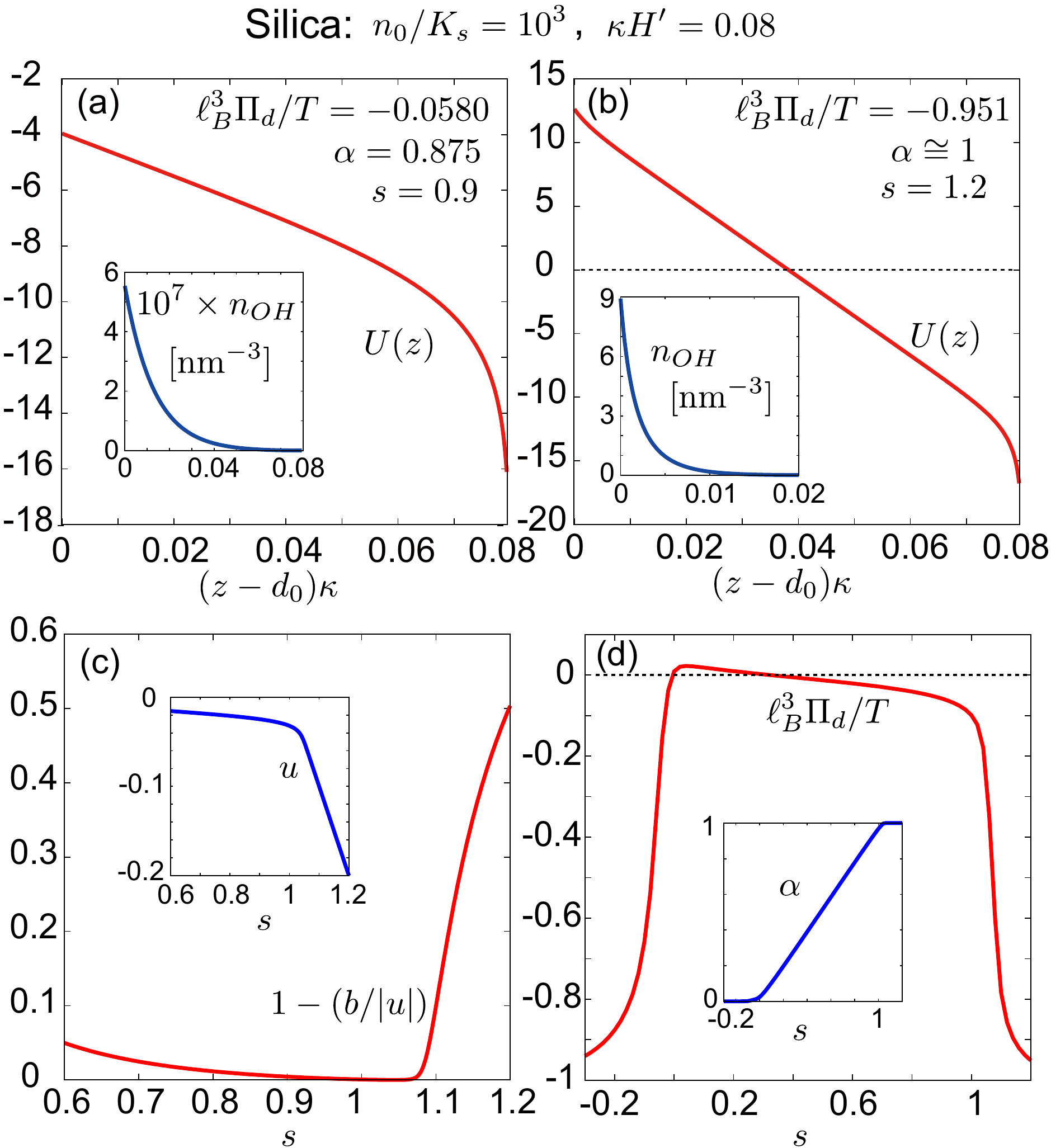}
\caption{ Results for silica with  
 $\kappa H'=0.08$  and $n_0/K_s=10^3$, where 
$\kappa=2.2\times 10^{-2}/$nm. 
Displayed is  $U(z)$ in the range $0<z-d_0<H'$ 
for   (a)  $s=0.9$, $\alpha= 0.875$,  and 
$\ell_B^3 \Pi_d/T=-0.0580$  
and  (b)  $s=1.2$,   $\alpha= 1.0$,  and 
$\ell_B^3 \Pi_d/T=-0.951$. 
Plotted also is $\nO(z)=\nO^0 e^{U(z)}$  near the bottom in units of nm$^{-3}$ 
(inset). 
(c) Screening fraction $1-b/|u|$ and $u$ (inset) vs  $s$.  
Screening is very weak for $s<1.1$.  
(d) $\ell_B^3 \Pi_d/T$ and $\alpha$  (inset) vs $s$, 
where $\Pi_d<0$ for most $s$. 
}
\end{figure}

In  Fig.15, we show typical 
profiles of $U$ for $u<0 $ and $s>0$ with $H'=5\ell_B =0.08/\kappa$, 
which exhibit     a  negative slope far from the walls.
Such profiles can appear for  $\Pi_d<0$, so we define 
a dimensionless number $b$ by 
\be 
 b= (-\Pi_d/2\pi T \ell_B )^{1/2}/\Gamma_0,
\en 
Then,   $\Pi_d/T\nO^0 = - 4A_1^2 b^2$. 
The slope of $U$ 
is  $ -2A_1 b \kappa$ and 
the corresponding electric field 
is   $4\pi e\Gamma_0 b/\ve_0$ far from the walls. 
Here, the  anions (OH$^-$) 
are accumulated at the bottom  
and the cations  (M$^+$ and H$^+$) 
at the top, but  their screening of the surface charges 
 is only partial.  
The accumulated ion numbers $N_i= \int dz n_i(z)$ satisfy  
\bea 
&&
N_{\rm OH}= \Gamma_0 (|u| - b),  \\
&&
N_H+N_M =  \Gamma_0  (s-b) , 
\ena
which are consistent with  Eq.(12). The degree of screening 
(screening fraction) 
is given by  $1- b/|u|$ at  the bottom and by   
$1-b/s$ at the top. It  is 0 with no screening (with no  diffuse layer)  
and is 1 for  complete screening.

In Fig.15, at $n_0/K_s=10^3$, 
we find $s=0.9$,  $\alpha=0.875$, and  $b=0.0245=37/A_1$ in (a) 
 and  $s=1.2$,  $\alpha=1.0$, and $b=0.099=150/A_1$  in (b).  Thus, 
   $1- b/|u|$ is 0.02 in (a) and 0.5 in (b). 
In fact,  $\nO(z)$ is very small  in (a) 
but is appreciable in (b) (inset). 
In (c), we show $1- b/|u|$ vs $s$ in the range $0.6< s<1.2$, 
which is very small  for $s<1.1$  but  grows abruptly for $s>1.1$. 
 For  $s<1.1$, there is almost no  screening  at 
the bottom and the screening fraction 
at the top is $ \alpha/s$.  
In (d), we plot $\Pi_d$ and $\alpha$ in the range $-0.2<s<1.2$, 
In accord with Fig.13, 
$\Pi_d$  is positive only in a narrow range of $s~ (\ls \alpha \sim 10^{-2})$.
 Here $u$ is negative and  $H'$ is small, but 
self-regulation behavior  $|u|\ll 1$   
in the range  $0<s< 1$ still persists. 

The region of  $\Pi_d>0$ is widened 
with increasing $n_0$.   See   the two panels 
in Fig.13, where $n_0/K_s=10$ and $10^3$.  
In Fig.16, this  is more apparent for $n_0/K_s=10^4$  
with  $H'=5\ell_B= 0.25\kappa^{-1}$, where   $\nM^0\cong \nO^0\gg \nH^0$.    
In (a), $ \Pi_d$  is about  $0.05T\ell_B^{-3}\sim 250T\nO^0$ for $0<s<1$, 
where $U(z)$ is negative  with a maximum about $-6.0$ and 
the ions in the cell are mostly M$^+$. 
However, $\Pi_d$ becomes negative abruptly 
for $s>1$. In (b), the profile of  $U(z) $ exhibits a changeover 
across  $s=1$  at this   $n_0$. 
For $s=0.4$, $E$ vanishes at the maximum point. 
However, once $\Pi_d<0$, 
 unscreening is significant 
at the bottom; in fact, $(u, 1-b/|u|)= 
(-0.012, 0.10)$ for $s=1.0$ 
and $(-0.10,  0.24)$ for  $s=1.1$.  
 The charcteristic features of these behaviors are unchanged 
even for  $n_0/K_s=10^5$.

The screeing  should be  easier for larger   $n_0$. 
This  is confirmed in  Fig.17, where  plotted is   
$1- b/|u|$ vs $n_0/K_s$ at  fixed $s$  
for (a) silica oxide surfaces 
  and (b) carboxyl-bearing surfaces. 
Here,   $H'$ is $ 5\ell_B$ in (a) and  is $50\ell_B$ in (b).  
The $K_s$ and   $\Gamma_0$ in the two systems 
are very different in  Eqs.(15) and (67). 
Nevertheless, we can see  significant unscreening    
at relatively small $n_0$  in the two systems. 
In addition, 
 the changeover from unscreening to  screening with increasing $n_0$  
is  nearly discontinuous for   $ 0<s<1$ 
(from $b\cong s-\alpha$ to $b\cong 0$),  
 while it   is gradual for 
$s>1$.

\begin{figure}[tbp]
\includegraphics[scale=0.41]{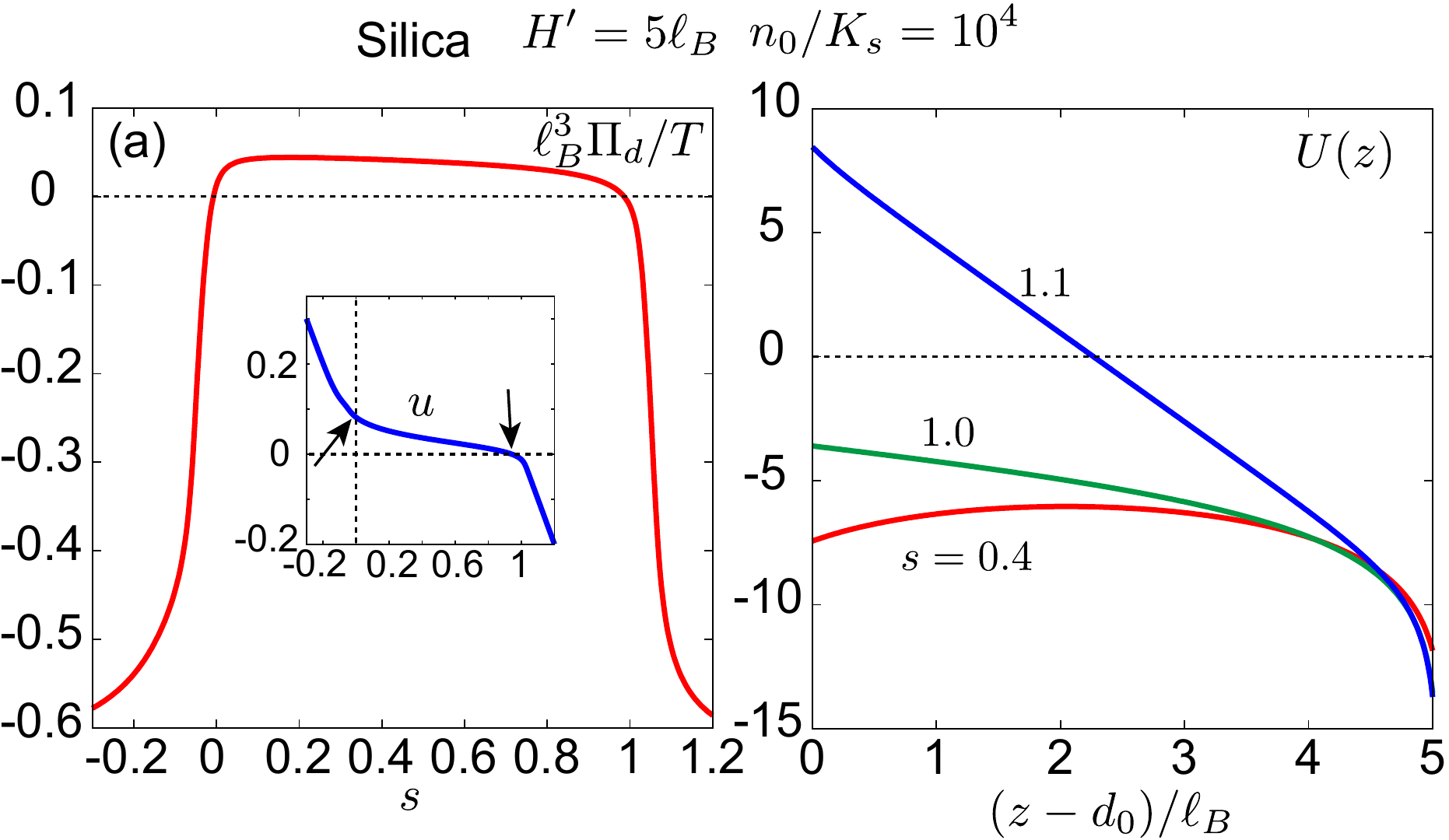}
\caption{ (a) $\ell_B^3\Pi_d/T$ vs $s$ 
in a cell with  $H'=5\ell_B= 0.25\kappa^{-1}$ 
for $n_0/K_s=10^4$.  For $0<s<1$, it is positive and is 
of order $0.05$ with small positive 
$u=\alpha-s$ (inset).   Point of $\Pi_d=0$ are marked by arrows. 
(b) Profiles of $U(z) $ 
for $s=0.4$, $1$, and $1.1$. 
}
\end{figure}

\begin{figure}[tbp]
\includegraphics[scale=0.41]{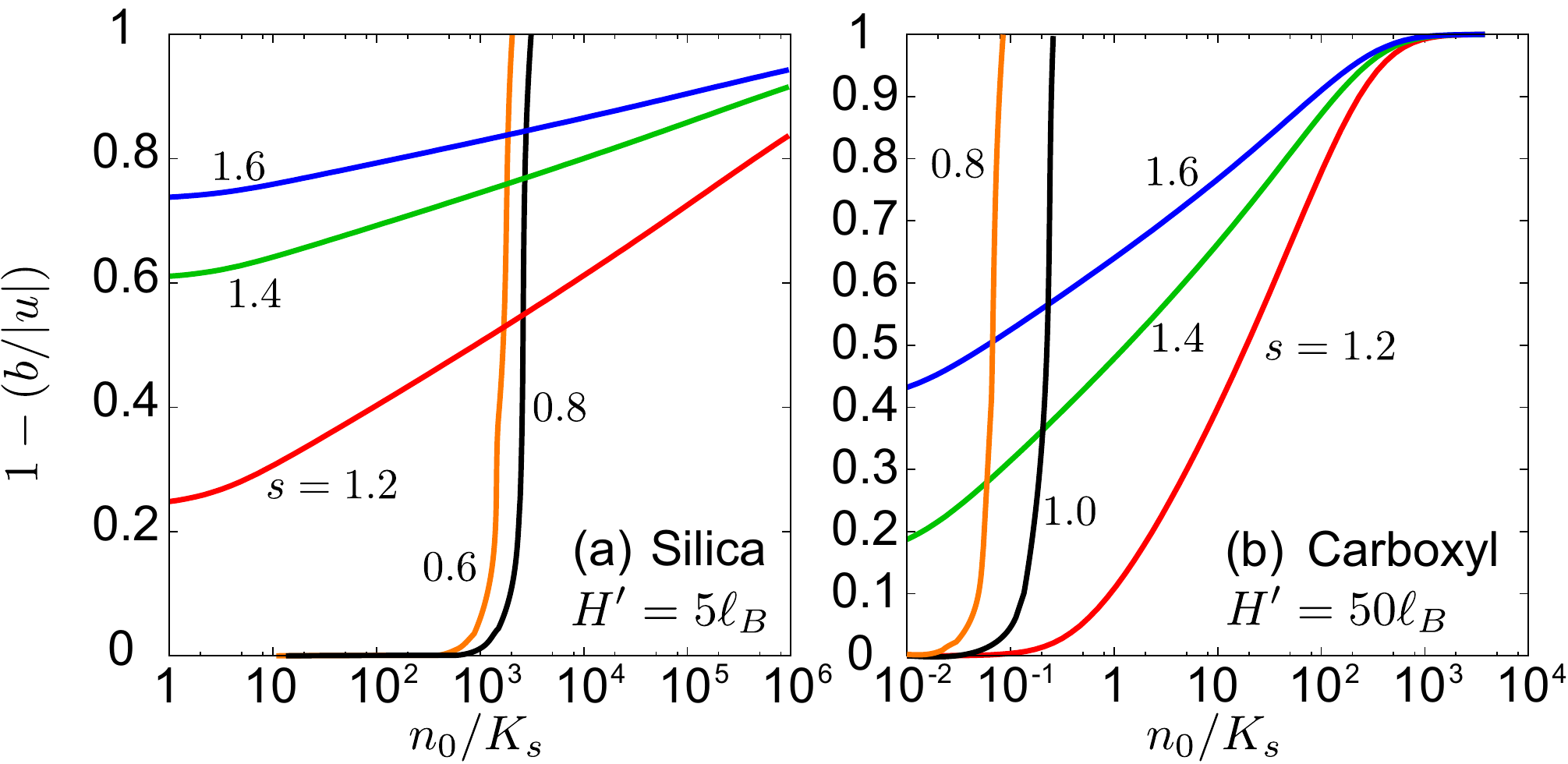}
\caption{  Screening fraction $1-b/|u|$ vs $n_0/K_s$ 
at the bottom 
for (a)  silica oxide surfaces with $H'=5\ell_B$ 
 and (b) carboxyl-bearing surfaces with $H'= 50\ell_B$ 
for five values of $s=\sigma_m/e\Gamma_0$. 
Screening increases with increasing $n_0$ and/or $s$. 
Changeover from unscreeing to screening 
is gradual for $s>1$ but is abrupt for $0<s<1$. 
}
\end{figure}

\subsection{Theory of partial screening in thin cells}

We finally present some analytic results on  the partial screening. See 
Appendix E for their derivations. With $b\gg A_1^{-1}$, 
we  introduce  a characteristic length $\ell_\infty$ by   
\be 
\ell_\infty = (2A_1 b)^{-1} \kappa^{-1}=(4\pi\ell_B \Gamma_0 b)^{-1}, 
\en  
which is of the same order as $\ell_{\rm GC}$ in Eq.(56) for $|u|\sim b$. 
If  we assume $U(H-d_H)=U_H<0$ and  $|U_H|\gg 1 $, 
 $U(z)$ near the top is well approximated  by  
\be 
U(z) 
\cong U_H + w +2\ln(1+a_H- a_H e^{-w})\quad ({\rm top}),
\en 
where $w= (H-d_H-z)/\ell_\infty$ and 
 $a_H= s/2b- 1/2$. The last term is zero at $w=0$ 
and tends to $2\ln(1+a_H)$ for $w\gg 1$,   so the profile 
changes on the scale of $\ell_\infty$. 
This formula  excellently describes the profiles in 
 (a) and (b) of Fig.15 and those in Fig.16b 
 near the top.  The  interior 
electric field is a constant  for  $\ell_\infty \ll H'$. 
For the profiles 
in  Fig.15a  and those of $s=1$ and 1.1 
in Fig.16b, we have 
$b\cong |u|$, so Eq.(83)  can be used even at  the bottom. 

For the profile in  Fig.15b, 
the diffusive layer near the bottom gives 
 $b\cong 0.5 |u|$. In this partial screening, 
 the profile at the bottom 
 is well approximated by  
\be
U(z)\cong
 { U}_0 -s -2\ln(1+a_0- a_0 e^{-s}) \quad ({\rm bottom}),
\en 
where  $U(d_0)=U_0\gg 1$,  
$s= (z-d_0)/\ell_\infty$,  and   $a_0= |u|/2b- 1/2$. 
For the profile in Fig.15b, 
 Eqs.(83) and (84) should  coincide far from the walls, so  we require 
\bea 
&&\hspace{-1.6cm} 
H'/\ell_\infty\cong U_0-U_H -2\ln [(1+a_0)(1+a_H)]\nonumber\\
&&\hspace{-9mm}\cong 
4\ln(4A_1b)+\ln \bigg[\frac{(|u|-b)(s-b)}{(|u|+b)(s+b)}\bigg]
\ena 
The right hand side can much exceed 1 for $A_1b\gg 1$.
 For (b) in Fig.15, Eq.(85) is  nearly exact.

\section{Summary and remarks} 

We have examined  ionization on a  dielectric film 
in water in applied  field.
 In  the geometry in Fig.1, the  surface 
 can be negatively charged with proton desorption. 
From the  mass action law, 
the degree of dissociation  $\alpha$  is   determined by  the ratio of 
the proton density     close to the film  $\nH(0)$ and 
the dissociation constant $K_s$($=10^{-{\rm pK}}$mol$/$L).
We have added  NaOH at a  density $n_0$ 
in water\cite{Yama} to decrease  $\nH(0)$. 
Main results  are summarized as follows.

(i) In Sec.II, we have presented the  free energy $F$, 
 depending  on the solute densities $n_i$ and the degree 
of ionization $\alpha$, where the the electrode charge density 
$\sigma_m$ or the potential difference $V$ is fixed. 
The contributions from  the  ionizable film and 
the Stern layers have been included. 
Minimization of the corresponding 
grand potential $\Omega$  yields equilibrium conditions 
including the mass action laws.  
The derivative of the equilibrium $\Omega$ 
 with respect to the cell width  $H$ 
yields    the osmotic pressure $\Pi$.

(ii) In Sec.III, we have assumed 
 $H \gg  \kappa^{-1}$. Analysis 
has been made on  silica oxide  films with pK$=7.3$ 
 in the nonlinear PB regime,  
where the Gouy-Chapman length $\ell_{\rm GC}$ 
 is shorter than $\kappa^{-1}$. 
We have obtained a simple equation for $\alpha$ 
in Eq.(61) for given $\sigma_m$ and  $n_0$.  As in Fig.7b, 
we have found remarkable  self-regulation 
behavior $\alpha\cong s= \sigma_m/e\Gamma_0$   
for  $0<s<1$ and 
$n_0\ll n_c$,  where $\Gamma_0$ is the areal density of  
the ionizable groups and 
$n_c$ is a crossover density. 
From Eq.(62), we find 
 $n_c= 10^{-2}$ mol$/$L 
 for silica oxide  surfaces. 

(iii) In Sec.IV, we have presented results using 
 the parameters of carboxyl-bearing surfaces with pK$=4.9$. 
General trends are 
common to those for slica oxcide surfaces as in Figs.8 and 9, 
but numerical factors are very different. 
For example, we have  $n_c= 2\times 10^{-5}$ mol$/$L.

(iv) In Sec.V, we have presented analysis for small 
$H'=H-(d_0+d_H) $  with a reservoir attached. 
 Without applied field,  the disjoining pressure $\Pi_d$ 
is proportional to  $ 
\sqrt{H'}$  for $H'<  \ell_{\rm GC}$ 
and  to $ (H')^{-2}$  
for $\ell_{\rm GC}<H'<\kappa^{-1}$.  In 
applied field,  the self-regulation  
($\alpha\cong s$ for $0<s<1$)    holds even 
for $\ell_B<H'<\kappa^{-1}$. We have also found that the surface charges can  
 be screened only partially  for not small $s$.  
For silica oxide srfaces,   $\Pi_d$ is mostly negative 
for  $n_0/K_s\ls 10^3$ in Figs.13 and 15(d),  but it assumes 
a large positive value  in the range $0<s<1$ 
for $n_0/K_s=10^4$   in Fig.16. Similar results follow  
for hydroxyl-bearing surfaces as in Fig.17. We have derived 
analytic expressions for the  potential profiles
in  partial screening.

(v) In Appendix A, we have examined  the experimental method 
of imposing the fixed charge condition. 
In Appendix C, we have  derived 
 the expressions for the osmotic and  disjoining 
 pressures.  In Appendix D, we have  derived 
the expressions for the surface free energy 
for ionizable surfaces in applied field.

We make some remarks.
 
(1) There are a variety of ionizable surfaces 
with very different parameters ($K_s$ and $\Gamma_0$)
\cite{Beh0,Beh1,Beh,Bie,BehR,Lowen} under strong influence 
of ions.  We have examined 
 dissociation  with small $K_s$, where the autoionization  of  water 
comes into play.  The  charge regulation 
has been  controlled by  the amount of  NaOH. 
If $K_s$  is much larger, 
 we may add  HCl to increase the bulk  
proton density $\nH^0$.  We can also add  KCl 
to increase $\kappa$, as discussed  below Eqs.(42), (46), and (74). 

(2) In future we should study  
dynamics of surface ionization in nonstationary electric field.
We note that the deprotonation  
 on a silica-water interface   takes  place very slowly 
as rare thermal activations\cite{Onukibook}. 

(3) There are a number of  nonequilibrium 
situations with chemical reactions\cite{Yama1,Bazant},
where  phase changes can take place.
For example, elelectrowetting\cite{Andel1,Mug} has been studied 
without chemical reactions.

\acknowledgments

This work was supported by KAKENHI No.25610122. RO
acknowledges support from the Grant-in-Aid for Scientific
Research on Innovative Areas Fluctuation and Structure (Grant No. 25103010)
from the Ministry of Education, Culture, Sports, Science, and
Technology of Japan.  We would like to thank Junpei Yamanaka 
for valuable discussions.

\vspace{2mm}
\noindent{\bf Appendix A: Realization of 
constant charge boundary condition }\\
\setcounter{equation}{0}
\renewcommand{\theequation}{A\arabic{equation}}
In Fig.1,  a  battery  has been used to produce 
 an equilibrium  electrolyte  state at a given potential difference $V$. 
If it is   disconnected, the surface charge density $\sigma_m$ 
(at the lower metal surface)  becomes fixed. 

As illustrated in Fig.1, we further  connect a small external condenser 
with a capacitance ${\cal C}_{\rm ex}$ to the electrodes. 
 The   potential difference of the capacitor is given by 
\be 
V_{\rm ex}=Q_{\rm ex}/{\cal C}_{\rm ex}
\en 
where  $Q_{\rm ex}$ is the initial charge. After this  connection, 
the surface charge density and the potential difference 
of  the electrodes  are changed as  $\sigma_m \to 
\sigma_m+\Delta\sigma_m$ and $V \to V+\Delta V$. Then, the 
   capacitor change is  changed by 
$- S_0 \Delta\sigma_m$, where $S_0$ is the surface area 
of the electrodes. If  the potential 
 equilibration is attained, we have  
\be 
V+\Delta V= [Q_{\rm ex}- S_0 \Delta\sigma_m]/{\cal C}_{\rm ex}
\en 
In the limit of small $\Delta\sigma_m$,  we may set   
 $\Delta V= \Delta\sigma_m /C_{\rm diff}$, where 
$C_{\rm diff}= \p \sigma_m/\p V$ is the 
differential capacitance of our system (per unit area). Then, Eq.(A2) yields 
\bea 
\Delta\sigma_m&=& (V_{\rm ex}-V)/[S_0 /{\cal C}_{\rm ex}+ 
1/ C_{\rm diff}] \nonumber\\
&& \cong (V_{\rm ex}-V){\cal C}_{\rm ex}/S_0,
\ena 
where the second line holds for $C_{\rm ex}\ll S_0 C_{\rm diff}$. 
Therefore, for sufficiently 
small ${\cal C}_{\rm ex}$, 
$\Delta\sigma_m$  depends only on the initial 
$V$ of our system and 
 remains fixed independently  of the subsequent 
 physical and chemical 
processes in the cell.

\vspace{2mm}
\noindent{\bf Appendix B: Bulk ion densities in terms of $\ns$}\\
\setcounter{equation}{0}
\renewcommand{\theequation}{B\arabic{equation}}

In Eq.(40)  we can assume the relation, 
\be 
 \nM^0+\nH^0 = \nO^0,
\en 
without loss of generality by 
 shifting the origin of $U(z)$ appropriately 
($U \to U+{\rm constant}$). Then, we obtain Eq.(41). 
From the chemical equilibrium 
conditions (4) and (5), $\nH^0$ and $\nM^0$ are expressed as 
\be 
n^0_{\rm H}= K_{\rm w}/n^0_{\rm OH}, \quad 
\nM^0= K_{\rm b}\ns/n^0_{\rm OH}.  
\en 
From Eqs.(B1) and (B2) $\nO^0$ is given by  
\be
 n^0_{\rm OH}=( K_{\rm w}+ K_{\rm b}\ns)^{1/2}.  
\en 
The density $n_0$ of M atoms in Eq.(7) is expressed as 
\be 
n_0= \ns (1+ K_{\rm b}/\nO^0).
\en 
For  $\kappa H\gg 1$,  $n_i^0$ 
coincide with those in Eqs.(6)-(8).

We can attach a reservoir with the same 
 $\ns$ as that in the cell. Then, the  reservoir 
densities of  H$^+$, OH$^{-}$, 
and M$^+$  are given by 
$n^0_{\rm H}$ and  $n^0_{\rm OH}$, and 
$n^0_{\rm M}$ in Eqs.(B2) and (B3). 
The reservoir  osmotic pressure  is written as      
\be 
\Pi_r^0 =T\sum_i n_i^0= T(\ns+ 2n^0_{\rm OH}).  
\en

\vspace{2mm}
\noindent{\bf Appendix C: Force between parallel walls  }\\
\setcounter{equation}{0}
\renewcommand{\theequation}{C\arabic{equation}}

We suppose  two  equilibrium states  
in the geometry in Fig.1, where  the cell length  is 
$H$ in one state and $H+\delta H$ in the other slightly 
elongated one. The  water density and the 
temperature are common. All the quantities 
are independent of the lateral coordinates $x$ and $y$. 
The  grand potentials  $\Omega$ and $\Omega'$ per unit area 
are defined in the two states from Eq.(32) or Eq.(35). 
We calculate    the difference 
$\delta\Omega=\Omega'-\Omega$ for small $\delta H$. 

In the elongated state,   the quantities are denoted with 
 a prime and  the space  coordinate is written as  $z'$, where 
 $n_i'(z')$ are 
 the  densities and $\alpha'$ is the degree of ionization. 
We impose   $\sigma_m'=\sigma_m$ at fixed charge 
and   $ V'=V$ at fixed potential difference. 
We assume a mapping relation
between the positions $z'$ and $z$  as \cite{Oka3,Onukibook} 
\be 
z' = z + w(z),
\en 
where   $w(z)$ is  a small {\it displacement}  with 
$w(0)=0$ and $w(H)=\delta H$. We consider the  deviations,   
\be
\delta n_i= n_i'(z')-n_i(z), \quad 
 \delta\alpha= \alpha'-\alpha.
\en

The mapping (C1) yields    $d/dz=(1+ w' )d/dz'$ and 
 $\int_0^{H+\delta H} dz' = \int_0^H dz (1+ w')$,
 with  $w'= dw/dz$. Then, from $\ve_0 dE'/dz'= 4\pi\rho'(z')$, 
 the   deviation $\delta E= E'(z')- E(z)$ is related to 
$\delta\rho= \rho'(z')-\rho(z)$  as 
\be 
\ve_0 \frac{d}{dz}\delta E=4\pi (w'\rho+ \delta\rho),
\en 
to linear order.  From  Eq.(23) the deviation  in the electrostatic free 
energy, $\delta F_e =F_e'-F_e$,    is calculated as     
\bea 
\delta F_e &=&
 \int dz w'[ \frac{\ve_0}{8\pi} E^2+ \rho\Phi] \nonumber\\
&+&  
\int dz \Phi\delta\rho + \Phi(0)\delta \sigma_{\rm A}+V\delta\sigma_m,
\ena 
where the first term arises from the cell elongation  
and the other terms  coincide with those in Eq.(24). 
Similarly, the elongation contribution to $F_b$ in Eq.(26) is 
given by $\int dz w' f_b$, where $f_b$ is the bulk free energy density 
in $F_b= \int dz f_b$.  From Eq,(32) or Eq.(35), we find  
\bea 
\delta\Omega &&= -\int dz w' \Pi - (N_{\rm M}+ N_{\rm MOH})
 \delta h_0\nonumber\\
&&=- \Pi \delta H - (N_{\rm M}+ N_{\rm MOH})  \delta h_0,  
\ena 
where $\delta h_0= h_0'-h_0$ and 
$\Pi$ is given in Eq.(44). 
 Since $\Pi$ is a constant, we obtain 
 the second line. 
Note that Eq.(C5) holds both  at fixed $\sigma_m$ 
and fixed  $V$. 

 The 1D  theory in this appendix 
can  be extended in various 3D situations. 
For example, we can  calculate the solvent-mediated interaction 
 between  colloidal particles in a 
mixture solvent \cite{Oka3}. 
The 3D mapping $x'_\alpha =x_\alpha + 
w_\alpha$   has been  used in  elasticity theory. 
Moreover, we can   use it to 
derive the stress tensor  
  for various fluids (including near-critical fluids,  
liquid crystals, and electrolytes)
\cite{Onukibook}.

\vspace{2mm}
\noindent{\bf Appendix D: Calculation of surface free energy}\\
\setcounter{equation}{0}
\renewcommand{\theequation}{D\arabic{equation}}

\begin{figure}[tbp]
\includegraphics[scale=0.4]{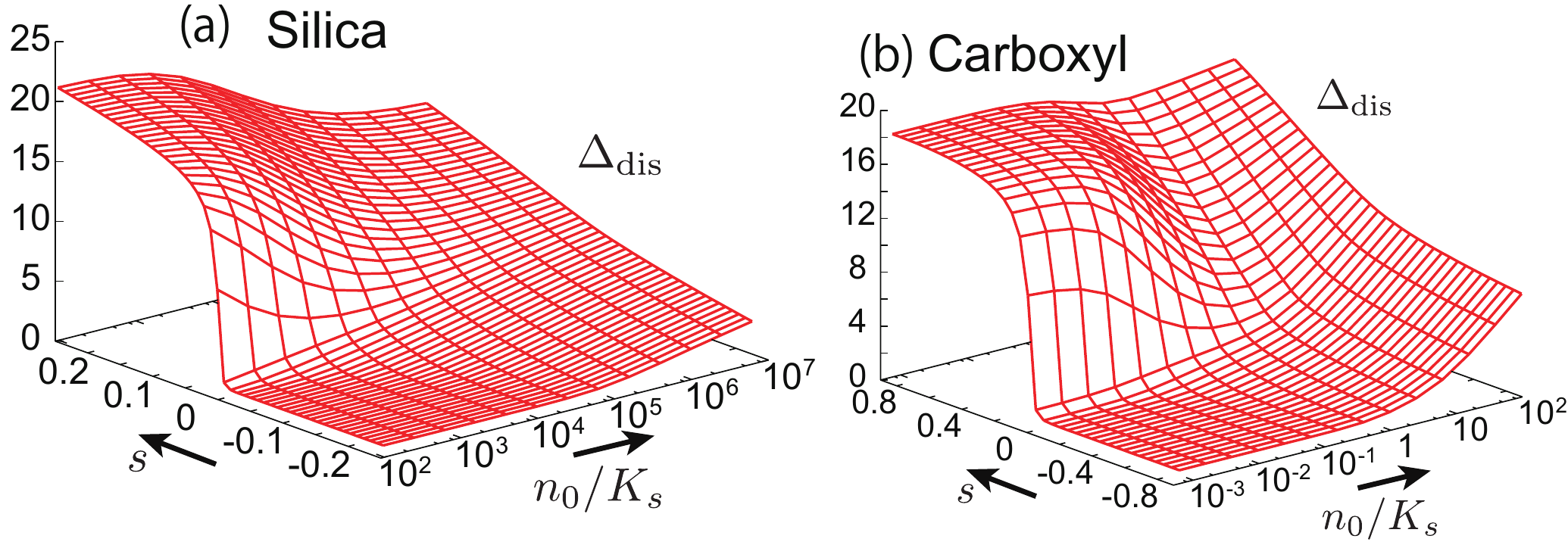}
\caption{ Normalized decrease in the dissociation free energy 
per ionized group 
$\Delta_{\rm dis} $ in Eq.(D5) 
  in the $s$-$n_0/K_s$ plane for (a) silica 
and (b) carboxyl-bearing interfaces. These are 
 calculated from the grand potential 
$\Omega$  in the limit $H\gg \kappa^{-1}$.  
}
\end{figure}

We calculate $\Omega$ in Eq.(35) at fixed 
$\sigma_m$ and $\nO^0$ using  Eqs.(18), (33), and (34). 
It is  rewritten  as 
\bea 
&&\hspace{-0.5cm} 
{\Omega}= -\int dz \bigg[T\sum_i n_i+ \frac{\ve_0 }{8\pi }E^2 \bigg] 
- \frac{\sigma_A^2}{2C_O } + \frac{\sigma_m^2}{2C_{\rm tot}} + {{F_s}}
  \nonumber\\
&&\hspace{-0.5cm}+T\Gamma_0 \alpha \ln[n_{\rm H}(0)\lambda_{\rm H}^3] 
  +[\Phi(d_0)-\Phi(H-d_H)]{\sigma_m}
\ena 
where  $C_{\rm tot}^{-1}= C_0^{-1}+ C_H^{-1}+C_d^{-1}$.  At fixed $V$, 
we should subtract $V\sigma_m$ from the right hand side. 

We assume  a  thick cell with 
 $\kappa H\gg 1$,  where $n_i \to n_i^0$ and  $E\to 0$  far from the walls. 
Then,  using   Eqs.(40) and (47), we can rewrite  Eq.(D1)  as  
\be 
\Omega= -H \Pi_r^0 + \gamma_H  +\gamma_0+ \sigma_m^2/2C_{\rm tot} .
\en 
where $\Pi_r^0$ is the reservoir osmotic 
pressure in Eq.(B5). The 
 $\gamma_H$ is the surface free energy from 
the upper diffuse layer (without surface ionization) 
given by  
\be
{\gamma_H}/T 
=  [1-({A_1^2s^2+1})^{1/2}]\kappa/\pi \ell_B + g(A_1s) \sigma_m/e ,
\en 
where $g(x)=2\ln[\sqrt{1+x^2}+x]$.
The $\gamma_0$ is that of  the lower surface 
and the difference $\Delta\Omega=\gamma_0-\gamma_H$ 
is the contribution due to the surface ionization of the form,  
\bea 
&&\hspace{-11mm}
\frac{\Delta \Omega}{T\Gamma_0} 
=\frac{2}{A_1} [(A_1^2s^2+1)^{1/2}-(A_1^2u^2+1)^{1/2}] + 
\frac{F_s}{T \Gamma_0 } 
\nonumber\\
&&\hspace{-8mm}
+ \alpha \ln[n_{\rm H}(0)\lambda_{\rm H}^3]
-s [g(A_1u)+g(A_1s)]  -\frac{A_2}{2}\alpha^2.
\ena

The above $\Delta\Omega$ 
coincides with $\Delta\Omega$ in Fig.5 calculated from 
integration of $d\Omega/d\alpha$ in 
 Eq.(36). In fact,  differentiation of  Eq.(D4) 
with respect to $\alpha$ at fixed $s$ yields  Eq.(36),  
since 
$\p [\ln[n_{\rm H}(0)]/\p\alpha= A_2 
+ 2A_1(1+A_1^2u^2)^{-1/2}$ from Eq.(59).  
In chemical equilibrium (11),  $\alpha$ and  $\Delta \Omega$ 
are  functions of $s$ and $n_0$, where  
 $F_s/T\Gamma_0 + \alpha \ln[n_{\rm H}(0)\lambda_{\rm H}^3]
=\ln (1-\alpha)=-\ln[1+ K_s/\nH(0)]$ in Eq.(D4). 
We define 
\be 
\Delta_{\rm dis}= -\Delta\Omega/T\Gamma_0\alpha,
\en 
in  equilibrium. 
Then, $-T\Delta_{\rm dis}$ 
 is the dissociation free energy per ionized group, 
 including   the effects  
of the  diffusive and Stern layers and the  electrode. 
In  Fig.18,  we  display  $\Delta_{\rm dis}$.  
For $n_0<n_c$, it  is close to 1 for $\alpha\ll 1$ (for  $s<0$)  
but it is  10-20  for not small $\alpha$  (for $s>0$).  
For $n_0>n_c$, it  increases gradually with increasing $s$.

\vspace{2mm}
\noindent{\bf Appendix E: Calculations for thin cells }\\
\setcounter{equation}{0}
\renewcommand{\theequation}{E\arabic{equation}}

{\it{Calculations for $s=0$}}.  
We first  present analysis in the relatively simple case 
of $s=0$ for $A_1\alpha \gg 1$, where  $U(z)<0$ 
with large $|U|$. 
Since $dU/dz$ is equal to  0 at $z=H-d_H$ and 
to  $2\kappa A_1\alpha$ at $z=d_0$,  Eq.(46) yields   
\bea 
&&\kappa^{-2} (dU/dz)^2\cong e^{-U(z)}- e^{-U_H},\\
&&  (2A_1\alpha)^2\cong e^{-U_0}- e^{-U_H},
\ena  
We  can integrate Eq.(E1) in the following form,  
\be 
\kappa (H-d_H -z)\cong {2} e^{ U_H/2} \tan^{-1} (\sqrt{q(z)} ) ,
\en 
where we use  
$\int_0^x dy (1/\sqrt{e^y-1}) 
=2 \tan^{-1}(\sqrt{e^x-1})$ by setting $y=U_H-U$.  We define  
\be 
q(z)= \exp[U_H- U (z)]-1.
\en 
where $q(H-d_H)=0$ and $q(d_0) = (2A_1\alpha )^2 \exp({U_H})$. 
There are two cases. If $q\ll 1$,  we can set $\tan^{-1} \sqrt{q}
 \cong (U_H- U)^{1/2}
\ll 1$  in Eq.(E3) to obtain $H'\ll \ell_{\rm GC}$ 
and  Eqs.(71)-(74).  
On the other hand, if $q\gg 1$, we can set $\tan^{-1} {\sqrt q} \cong\pi/2$ 
to obtain $H'\gg \ell_{\rm GC}$ and  Eqs.(76)-(78). 

{\it{Calculations of partial screening}}.  
We explain the partial screening in Fig.15 
 with $s>0$ and $u<0$.   In terms of $b$ in Eq.(79),  
the PB equation in  Eq.(46) becomes   
\be
 2\cosh (U)-2 = 
\kappa^{-2} (dU/dz)^2-  (2A_1b)^2 .
\en 
 At $z=d_0$ and $H-d_H$, we find 
\bea  
&&\cosh(U_0)-1=2A_1^2( u^2-b^2), \\
&&\cosh(U_H)-1=2A_1^2( s^2-b^2), 
\ena   
If $U_H<0$ and $|U_H|\gg 1$, 
we can replace $\cosh (U)-1$  by 
$e^{-U}/2$  in Eq.(E5) near the top. 
It  follows Eq.(83). Integration of 
$\nH(z)/\nH^0=\nM(z)/\nM^0= e^{-U}$ near the top 
can be performed to give  Eq.(81). On the other hand, 
if  $U_0\gg 1$, we can replace $\cosh (U)-1$  by 
$e^{U}/2$ near the bottom. Then, Eq.(84) is obtained 
and  integration of $\nO(z)/\nO^0= e^{U}$ near the bottom 
 yields Eq.(80).


\begin{thebibliography}{0}


\bibitem{Is} J. N. Israelachvili,  
{\it Intermolecular and Surface 
Forces} (Academic Press, London, 1991). 


\bibitem{Hunter} R. J. Hunter, Foundations of Colloid Science 
(Oxford University Press, Oxford, 2001). See Chap.10 in this book.

\bibitem{Butt} 
H.-J. Butt, K. Graf, and M. Kappl, Physics and Chemistry of
Interfaces, 3rd ed. (Wiley-VCH Verlag GmbH, Weinheim, 2013).


\bibitem{Andel3}
D. Ben-Yaakov,  D. Andelman,  R. Podgornik, D.  Harries, 
Current Opinion in Colloid and  Interface Science {\bf 16}, 542 (2011). 


\bibitem{Fuer} 
 T. W. Healy and D. W. Fuerstenau, J. Colloid Sci. {\bf 20}, 376 (1965).

\bibitem{Wes} 
J. Westall and H. Hohl, Adv. Colloid Interface Sci. {\bf 12}, 265 (1980).
%


\bibitem{Hie} 
T. Hiemstra, J. C. M. de Wit, and W. H. van Riemsdijk, J. Colloid Interface
Sci. {\bf 133}, 105 (1989).

\bibitem{Leckie} 
K. F. Hayes, G. Redden, W. Ela, and J. O. Leckie, 
J. Colloid and Interf. Sci. {\bf 142}, 448 (1991).
\bibitem{Sonn} J. J. Sonnefeld, J. Colloid Interface Sci. {\bf 155}, 191 (1993).


\bibitem{Hal} R.E.G. van Hal, J.C.T. Eijkel, and 
P. Bergveld,   Adv. Colloid Interface Sci.{\bf 69},  31 (1996). 






\bibitem{Is1} 
J. N. Israelachvili and  R. M. Pashley,
J.  of Colloid and Interface Science,  {\bf 98}, 500 (1984). 

\bibitem{Ga} A. Grabbe,  Langmuir  {\bf 9}, 797 (1993).

\bibitem{Zhao} 
C. Zhao, D. Ebeling, I. Siretanu, D. van den Ende,  and F.  Mugele,
 Nanoscale {\bf 7}, 16298 (2015).


\bibitem{BehR} G. Trefalt, S. H. Behrens, and M. Borkovec, 
Langmuir  {\bf 32}, 380 (2016).


\bibitem{Par} B. W. Ninham and V. A. Parsegian, 
J. Theor. Biol. {\bf 31}, 405428 (1971). 

\bibitem{Chan} 
 D. Y. C. Chan, T. W. Healy, and L. R. White, 
J. Chem. Soc., Faraday Trans. I, {\bf  172}, 2844 (1976). 



\bibitem{Beh0} 
S. H. Behrens and M. Borkovec, J. Phys. Chem. B {\bf 103}, 2918 (1999); 
Phys. Rev. E {\bf 60}, 7040 (1999).
\bibitem{Beh1} 
S. H. Behrens, D. Iso Christl,  R. Emmerzael, P. 
 Schurtenberger,  and M.  Borkovec
Langmuir {\bf 16}, 2566 (2000).
\bibitem{Beh}
S. H. Behrens and D.G. Grier, J. Chem. Phys. {\bf 2001}, {\bf 115}, 6716 (2001). 
\bibitem{Bie} 
P. W. Biesheuvel and  W. B. S. de Lint, 
 J. Colloid Interface Sci.  {\bf 241}, 422 (2001); 
P. M.  Biesheuvel, Langmuir  {\bf 17}, 3553 (2001). 


\bibitem{Lowen} 
M. Heinen, T. Palberg,  and H. L$\ddot{\rm o}$wen
J. Chem. Phys. {\bf 140}, 124904 (2014)



\bibitem{AndelEPL} 
T. Markovich, D. Andelman, and R. Podgornik, 
EPL, {\bf 113} ,26004 (2016). 
Their expression for $\Pi_d$ in the Ninham-Parsegian 
regime is obtained if $\nO^0/\nH^0= 
(\nO^0)^2/K_{\rm w}$ in our expression in 
Eq.(73) is replaced by 1.

\bibitem{Eigen} 
M. Eigen and L. De Maeyer, Z. Elektrochem. {\bf 59}, 986 (1955); 
M. Eigen and L. De Maeyer, Proc. Roy. Soc. (London), Ser. A {\bf 247}, 
505 (1958).

\bibitem{Gisler} 
T.  Gisler, S. F. Schulz, M. Borkovec, H. Sticher, P. Schurtenberger, 
B. D'Aguanno, and R. Klein, J. Chem. Phys. {\bf 101}, 9924 (1994).

\bibitem{Beh4}
R. Pericet-Camara,  G. Papastavrou, S. H. Behrens, 
 and M. Borkovec, J. Phys. Chem. B {\bf 108}, 19467 (2004). 


\bibitem{Yama} J. Yamanaka, Y. Hayashi, N. Ise, and T. Yamaguchi, 
 Phys. Rev. E {\bf 55},  3028 (1997). 

\bibitem{Yama1} 
M. Murai, H. Yamada, J. Yamanaka, S. Onda, 
M.  Yonese,   K. Ito,  T.  Sawada,  F.  Uchida,  and Y. Ohki, 
Langmuir {\bf  23}, 7510 (2007). 


\bibitem{Andelbook} D. Andelman, 
{\it  Introduction to Electrostatics in Soft and
Biological Matter.} 
 In Soft Condensed Matter Physics in Molecular and
Cell Biology; Poon, W., Andelman, D., Eds.; Scottish Graduate Series;
Taylor $\&$  Francis: New York, 2006; p 97.  


\bibitem{OnukiReview}
  A. Onuki,  R. Okamoto, and T. Araki, 
Bull. Chem. Soc. Jpn. {\bf 84} (2011) 284113.


\bibitem{Oka2}
R. Okamoto and A. Onuki, Phys. Rev. E {\bf 84}, 051401 (2011)



\bibitem{Raphael} 
 E. Raphael and J. F. Joanny, Europhys. Lett. {\bf 13}, 623 (1990).

\bibitem{Bor}  I. Borukhov, D. Andelman, and H. Orland, 
Eur. Phys. J. B {\bf 5}, 869 (1998). 


\bibitem{Netz} 
Y. Burak and R. R. Netz, J. Phys. Chem. B {\bf 108}, 4840 (2004)

\bibitem{Oka0} 
R. Okamoto and A. Onuki, J. Chem. Phys. {\bf 131}, 094905 (2009).

\bibitem{Oka1}
A. Onuki and R. Okamoto, J. Phys. Chem. B {\bf 113}, 3988 (2009).


\bibitem{Muth} M. Muthukumar, J. Chem. Phys. {\bf 120}, 9343 (2004); 
M. Muthukumar,  J. Hua, and A. Kundagrami, J. Chem. Phys.  
{\bf 132}, 084901 (2010). 


\bibitem{Olivera} G. S. Longo,  M. O. de la Cruz,  and I. Szleifer,
Soft Matter, {\bf  8}, 1344 (2012). 

\bibitem{Araki} T. Araki,  Soft Matter, {\bf 12}, 6111 (2016).

\bibitem{Zimm} C. B. Post and B. H. Zimm, Biopolymers 21, 2139 (1982); .
P. G. Arscott, C. Ma, J. R. Wenner, and V. A. Bloomfield, Biopolymers
36, 345 (1995);  A. Hultgren and D. C. Rau, Biochemistry 43, 8272 (2004); 
 C. Stanley and D. C. Rauy, Biophys. J. 91, 912 (2006).


\bibitem{Kolb} 
D. M. Kolb, Surf. Sci., {\bf  500}, 722 (2002).

\bibitem{Hautman} J. Hautman, 
J. W. Halley, and Y.-J. Rhee, J. Chem. Phys. 91, 467 (1989).


\bibitem{Yeh} 
 I.-C. Yeh and M.L.  Berkowitz,  J. Chem. Phys.  111, 3155 (1999).

\bibitem{Hender} P. S. Crozier, R. L. Rowley, and D.  Henderson, 
J. Chem. Phys.  {\bf 113}, 9202 (2000).

\bibitem{Madden} 
A. P.Willard, S. K. Reed, P. A. Madden, and D. Chandler, Faraday Discuss.
{\bf 141}, 423 (2009).

\bibitem{Takae} 
K. Takae and A. Onuki, J. Phys. Chem. B {\bf 119}, 9377 (2015); 
J. Chem. Phys. {\bf 143}, 154503 (2015)


\bibitem{comment1} 
As the Stern potential drop, 
we find the expression 
 $V^{\rm S}_0=4\pi\int_0^{d_0}dz [P_b-P(z)]$, where 
$P_b=(\ve_0-1)E(d_0)/4\pi$ is the bulk polarization close to the 
bottom Stern layer \cite{Takae}.


\bibitem{Ohshima} 
H. Ohshima, {\it Theory of Colloid and Interfacial Electric
Phenomena} (Elsevier, New York, 2006).


\bibitem{Landau}
L. D. Landau and E. M. Lifshitz, Electrodynamics of Continuous
Media (Pergamon, New York, 1984), Vol. 8.


\bibitem{Andel1} D. Klarman and D. Andelman, Langmuir {\bf 27}, 6031 (2011).
\bibitem{Mug} F. Mugele and J.C.  Baret,  J. Phys.: Cond. Mat. {\bf 17}, 
R705  (2005).


\bibitem{Oka3}
R. Okamoto and A. Onuki, Phys. Rev. E {\bf 88}, 022309  (2013). 




\bibitem{Aoki} 
K. Aoki, T. Li, J. Chen, and T. Nishiumi, J. Electroanal. Chem. {\bf 
613},  1 (2008); 
K. Aoki, T. Li, J. Chen, and T. Nishiumi, 
 J. Electroanal. Chem. {\bf 633}, 319 (2009). 


\bibitem{comment2} Setting $\Pi_d=0$, we solve 
Eq.(41) for  $u<0$ and $s>0$ exactly 
as $\tanh(U/4)=  -C \exp[\kappa (z-d_0)]$.  
   For  $A_1|u|\gg 1$ we have $C\cong 1+ 1/A_1|u| $ 
and  $s(|u|-\alpha) \cong \alpha/2A_1\kappa H'$. 
Then, $s\cong \alpha$ for $ H'>1/ A_1\alpha\kappa=\ell_{\rm GC}$.



  
\bibitem{Onukibook}  A. Onuki,  {\it Phase Transition Dynamics} 
(Cambridge University Press, Cambridge, 2002). See Appendix 6A in this book. 


\bibitem{Bazant} 
M. Z. Bazant, M. S. Kilic, B. D. Storey, and A. Ajdari, Adv. Colloid Interface
Sci. {\bf 152}, 48 (2009).

\end{thebibliography}
\end{document}